\documentclass[12pt]{article}
\usepackage{bm}
\usepackage{a4wide,graphicx}
\usepackage{cite}
\begin{document}

%\tightenlines

\title{$D$ Mesons in Nuclear Matter: A $DN$ Coupled-Channel Equations Approach}

%\author{A. Ramos$^1$, and T. Mizutani$^{1,2}$}
%\address{$^1$ Departament d'Estructura i Constituents de la Mat\`eria,
%Universitat de Barcelona, \\
%Diagonal 647, 08028 Barcelona, Spain \\
%$^2$ Department of Physics, Virginia Polytechnic Institute and State University \\
%Blacksburg, VA 24061, USA}

\author{T. Mizutani$^{1,2}$ and A. Ramos$^1$\\
$^1$ Departament d'Estructura i Constituents de la Mat\`eria\\
Universitat de Barcelona,
Diagonal 647, 08028 Barcelona, Spain \\
$^2$ Department of Physics, Virginia Polytechnic Institute and State
University \\
Blacksburg, VA 24061, USA}

\date{\today}

\maketitle
\begin{abstract}

A  set of coupled two-body scattering equations is  solved for the $DN$ system
embedded in an iso-symmetric nuclear matter. The in-medium behavior of charmed
$D$ mesons: $(D^+,D^0)$, is investigated from the self-consistent solution
within this scheme.  The effective meson-baryon Lagrangian in charm quantum
number one sector, the key ingredient in the present study, is adopted  from a
recent model by Hofmann and Lutz which has aimed at combining the charmed 
meson degree of freedom in a consistent manner with chiral unitary models.
After a critical examination, the original model is modified in several
important aspects, such as the method of regularization, in order to be more 
consistent and practical for our objective. The resultant interaction is  used
to reproduce the position and width of the $s$-wave $\Lambda_c(2593)$ resonance
in the isospin zero $DN$ channel. In the isospin one channel, it generates a
rather wide resonance at $\sim 2770$ MeV. The corresponding in-medium solution
is then sought by incorporating Pauli blocking and the  $D$- and
$\pi$-meson  dressing self-consistently. At normal nuclear matter density, the
resultant $\Lambda_c (2593)$ is found to stay narrow and shifted at a lower
energy, while the $I=1$ resonance is lowered in position as well and
broadened considerably. 

\end{abstract}
\vskip 0.5 cm



%\tightenlines

\section{Introduction}
\label{sec:intro}

The present article is devoted to the
study on the behavior of the $D$-meson in a cold symmetric (viz. total isospin
zero) nuclear matter by employing a set of  in-medium $DN$ coupled channel
equations to be solved self-consistently.

Of numerous theoretical investigations to date on the properties of mesons in
nuclear medium, one may notice a recent interest directed towards the in-medium
behavior of the (open charm) $D$ mesons in studying the existing, anticipated,
or speculative experimental outcome in relativistic heavy ion collisions,
antiproton reaction with nuclei, possible $D$-nuclear bound states, etc.
\cite{HIG00,MOR02,TSU99, SIB99,  MIS104, TOL04, LUT05}.  The primary
theoretical effort in this regard has been to understand how the mass of the
$D$ meson gets modified in nuclear matter: either normal or hot and/or dense.
The main objective is to know, for example, if the $D$ meson mass is reduced
significantly in a medium formed by heavy ion collisions. If this were the
case, then that could lead to an enhanced $D$  production during the
processes, bringing  a  possible conventional hadron physics scenario for the
suppression of the $J/\Psi$ production, often attributed to the long-time
speculated and more exotic process of the formation of the Quark-Gluon  plasma
due to deconfinement. In the present work we shall adopt a self-consistent
many-body coupled-channel method based  on a hadronic effective Lagrangian
which has enjoyed its success in studying the physics of $\bar K$ and $K$
mesons  in nuclear matter. As a matter of fact the methods of study on this
subject as employed in the above publications  follow rather closely the ones
as applied to the study of these mesons in nuclear medium which was initially
triggered by the issue of possible kaon condensate \cite{KAP86, NEL87}. They
are (i) the QCD sum rule method; {\bf QCDSR},  (ii) the nuclear mean field
approach; {\bf NMFA},  and (iii) the so-called self-consistent coupled
channels  method;  {\bf SCCM}. On the other hand there are  approaches based
upon effective quark potentials such as in   Refs.\cite{DIG01} and \cite{FRI02}
which are in a way complementary to these three approaches. We shall not
discuss those quark-model methods here. 

 The organization of the present article goes as follows. In
Sect.~\ref{sec:retro} we have a critical retrospect of the  related works
within the three methods stated above in order to motivate the present one.
Sect.~\ref{sec:review} is devoted to critically reviewing a series of works:
\cite{LUT05} and \cite{HOF05}, which had  motivations close to ours,  then
explain why we have come to adopt a somewhat different method by modifying what was 
used in those works. Sect.~\ref{sec:free}  presents the results of  our study 
of the $DN$ interaction obtained from a coupled-channels equation in free space. The
implementation of various medium effects on the properties of the $D$ meson
in nuclear matter is discussed in Sect.~\ref{sec:medium} and our
results are presented in Sect.~\ref{sec:results}.  
Sect.~\ref{sec:conclusion} is devoted to our conclusion and final remarks.
Those who are familiar with
the subject might skip some parts of the next section.

\section{Critical Retrospect}
\label{sec:retro}

In what follows, we shall outline  each of the methods mentioned above: {\bf
QCDSR, NMFA} and {\bf SCCM}, as the basis of why  we are motivated to take the
steps presented  in this work.   We believe this to be appropriate 
because, to date, no such account has been given to compare
different  approaches.  For this objective  our subsequent discussion will be heavily
inclined to review  the approaches employed in the physics of $\bar K$ in
nuclear matter  because the methods listed above for the $D$  meson have been
extensively used for the former. This is due to the apparent similarity
between the  $\bar K$ and $D$ \cite{TOL04}
as discussed later in the section.

To begin with we note that in all these three approaches (except for Refs.
\cite{TSU99, SIB99} to be  touched  upon later) the central entity is the meson 
propagator (or the
correlator) in the nuclear environment, $< T\left [D(x)\bar D(y)\right ]>_{nuc}$,
where $D(x)$ is the $D$-meson interpolating field.

\subsection{QCDSR}

The method \cite{HIG00, MOR02} exploits the quark-hadron duality to calculate the $D$ meson 
propagator in two different ways.  On the one hand it is written in terms of the underlying quark 
fields. Then by means of
short distance operator product expansion (OPE)
within perturbative QCD,  it is expressed in terms of basic QCD constants and condensates.  
On the other hand, within the hadronic picture
it is expressed in a spectral representation with a few adjustable parameters.
Then the two sides are matched in deep Euclidean region to extract hadronic quantities.
In the OPE expression
it is shown \cite{MOR02} that the essential  ingredient  is the product of the charm quark mass: $m_c$, and
the {\it in-medium}
light scalar $\bar qq$ condensate: $<\bar qq>_{nuc}$.  Following Ref.~\cite{COH92}, 
 the in-medium $\bar qq$ condensate is approximated by a sum of the vacuum part and the in-medium
correction, the latter being  the product of nuclear density and the nucleon matrix element of 
$\bar qq$,

\begin{equation}
<\bar qq>_{nuc}\approx <0|\bar qq|0> + \rho_N <N|\bar qq|N> \ .
\label{eq:condensate}
\end{equation}
On the other hand, by a factorization ansatz combined with  a  linear
density approximation \cite{DRU90} valid for low nuclear density,   the
in-medium $D$ propagator is written as a sum of the  free space part and the
in-medium correction which is proportional to the nuclear density $\rho_N$
times the free space $DN$ scattering amplitude, which is further approximated 
by the  $DN$ scattering
length.  Then QCDSR is eventually used to express the free space $DN$
scattering length in terms of a few QCD parameters, $m_c<N|\bar qq|N>$ in particular \cite{HIG00}. 
Finally, by
matching, the in-medium  meson mass shift is obtained as proportional to  the
nuclear density times the thus obtained $DN$ scattering length.
This relation is naturally expected from optical models in standard scattering
theory in which one obtains the in-medium meson self-energy as  the product (or
convolution) of  nuclear density  and the meson-nucleon scattering amplitude.
 Similar quantities appear in NMFA and SCCM
discussed later.  In the present approach the medium effect enters
through the linear dependence in $\rho_N$ only. At normal nuclear matter 
density the isospin averaged 
$D$  mass shift is obtained as  $\approx -50$ MeV. In
Ref.~\cite{MOR02} an additional mass shift due to  the time component of the in-medium vector $\bar qq$ 
condensate is reported as  $\sim \pm 25$ MeV for $\bar D (D)$.  A rather
strong sensitivity to the assumed high energy behavior of the spectral function
is noted in  \cite{MOR02}, which may be related to the difficulty in
determining the free space $D$ mass in QCDSR \cite{REI85}.  So along with various approximations 
mentioned earlier, results mentioned here should be regarded as semi-quantitative.  However, an
important finding is the large (attractive) contribution
induced by $\approx m_c<N|\bar qq|N>$ from OPE.  This term,  which is of
scalar-isoscalar in nature, enters just like the familiar pion-nucleon $\sigma$
term. But it is at least two  orders of magnitude larger because the charm
quark mass: $m_c\approx 1400$ MeV, is multiplied instead of the average
light quark mass. 
%One thus obtains a hint that a strong scalar-isoscalar attraction
%of some type may be present in the $DN$ interaction.

\subsection{NMFA}

We first outline  prototype of this approach used in the first
kaon condensation study in Refs.~\cite{KAP86, NEL87}.  There, the leading term in
 the  Lagrangian is from the non-linear realization of
chiral symmetry for the interaction of  Goldstone bosons with octet
baryons. This is 
supplemented by  symmetry breaking terms linear in the quark mass matrix.
 With  $N$ and $K$ being  the nucleon and kaon fields
respectively, the resulting kaon-nucleon interaction reads,
\begin{equation}
{\cal L}_{KN} = -i\frac{3}{8 f^2} \bar N\gamma^{\mu}N \bar K
\stackrel{\leftrightarrow}{\partial_{\mu}}K
+\frac{\Sigma_{KN}}{f^2}\bar N N \bar KK    \  ,
\label{eq:KN}
\end{equation}
where $\bar K \stackrel{\leftrightarrow}{\partial_{\mu}}K \equiv \bar K
\partial_{\mu} K - (\partial_{\mu} \bar K) K$ and $f$ is the Goldstone boson
decay constant. The first term is the Tomozawa-Weinberg (T-W)  vector
interaction. The second term provides a scalar-isoscalar  attraction 
characterized by  $\Sigma_{KN}$. This quantity, called the 
$\Sigma$-term, is  expressed by three $low\ energy \ constants$ that  may be written 
in terms of the  $\pi N$ and $KN$
$\sigma$-terms: $\sigma_{\pi}$, and $\sigma^{(i)}_K, \ (i=1,2)$  which are the
measures of chiral symmetry breaking.  
In NMFA, the meson (here it is the kaon) self-energy, $\Pi_K(p_0, \vec p\, )$,
which provides the dispersion equation relating
the energy ($p_0$) and  momentum ($\vec p$\,), 
%The inverse of the momentum space meson
%(here it is the kaon) propagator in nuclear medium reads,
%\begin{equation}
%D_K^{-1}(p_0, \vec p \, ) =p_0^2-{\vec p\,}^{2} -m_{K}^2-\Pi_K(p_0, \vec p\, )\ ,
%\label{eq:invprop}
%\end{equation}
%where $\Pi_K(p_0, \vec p\, )$ is the in-medium kaon self-energy.  Setting this
%inverse propagator  equal to zero provides the dispersion equation relating
%the energy ($p_0$) and   momentum ($\vec p$\,). 
%In NMFA,  $\Pi_K(p_0, \vec p\,
%)$ 
is  density times the nuclear expectation value of the above
meson-nucleon interaction: $-\rho_N< {\cal L}_{KN}>_{nuc}$.  With a simple Fermi gas
model for an isosymmetric nuclear matter, the energy of the kaon at rest is
obtained as (notice the difference between the in-medium kaon and anti-kaon due to the T-W vector interaction),
\begin{equation}
p_{0{(K, \bar K)}}= \sqrt{(m_K^s)^2+ (\frac{3\rho_0}
{8 f^2})^2} \pm \frac{3\rho_0}{8 f^2} \  ,
\label{eq:Kdisp}
\end{equation}
where $(m_K^s)^2 = m_K^2-\rho_s \Sigma_{KN}/f^2$ is the  square of the
in-medium (scalar) kaon mass with $\rho_s$ being the nuclear scalar density, 
and  the ordinary nuclear matter (vector) density is $\rho_0 \equiv \rho_N$.
In the non-relativistic limit, $\rho_0=\rho_s$. From the
above result, a considerable reduction in the kaon mass (more precisely the
energy at zero momentum) in a high density medium might be expected (for both
$K$ and $\bar K$) if the strength $\Sigma_{KN}$  becomes sizable, leading 
even to a possible kaon condensation.  
Further investigation on the $K$ mesons in hot and/or dense nuclear medium
with refinement may be found, for example, in
\cite{MAO99, MIS204}. See Ref.~\cite{MAR05} for an extensive set of references.

The above method has been  extended to the study of the $D (\bar D)$ mesons in an isospin
symmetric nuclear  matter in Ref. \cite{MIS104}. Two steps are required to reach the
goal. First one is to describe the static nuclear properties, viz. binding
energy per nucleon, compressibility, etc. in a  mean-field approach imposing
$SU(3)$ symmetry  with the non-linear realization of  chiral
symmetry  \cite{PAP99}. It is an extention of the original Walecka
``$\sigma$-$\omega$'' model \cite{SER86}. The optimal solution is found by imposing a stationary
condition on the free energy of the system with respect to the variation of the mean scalar and vector 
meson fields for a given value of nuclear density.  The 
meson-nucleon coupling constants are fixed at the normal nuclear saturation density.  
The second step
is to construct an {\it in-medium} $D$ meson-$nucleon$  interaction. 
The approach used for the in-medium kaon is extended to incorporate the
$D$ meson by using  gauged $SU(4)$ vector mesons, a method adopted in part from
\cite{LIU02}. The outcome is a  Lagrangian  similar to Eq.~(\ref{eq:KN})  in 
Refs. \cite{KAP86, NEL87}, but with a few  additional  terms, (see
Ref.\cite{MIS204} for the case of kaons in matter). Of particular interest are
the ones describing the interaction of the $D$ with  {\it mean} scalar
($\sigma$) and vector ($\omega$) meson fields.  They are made to contribute
to the in-medium $D$ meson interaction Lagrangian as
\begin{equation}
 g_{D\sigma}\bar DD\sigma  -ig_{D\omega}\bar D
\stackrel{\leftrightarrow}{\partial_{\mu}}D \omega^{\mu}
\ ,
\label{eq:meanfields}
\end{equation}
where the mean density dependent scalar and vector meson fields
% are from the
%optimization  condition for nuclei (or nuclear matter) in the {\it first step}
%mentioned above. So they 
are complicated
functions of nuclear density, effective (mean field) meson-nucleon coupling
constants, etc. 
%Otherwise there is no explicit appearance of nucleon variables
%in the interaction. 
Contributions arising from those $D$ interactions with
in-medium mesons contribute to the meson self-energy.
% expression, recall
%Eq.~(\ref{eq:invprop}). So 
For example, the mean $\omega$ field contribution
$g_{D\omega}\omega^0$ is to be added to the  one from the  T-W 
interaction, viz. the  $3\rho_0/(8 f^2)$ term in Eq.~(\ref{eq:Kdisp}).
% for
%the kaon.  Since mean meson fields carry density dependence, the in-medium $D$ meson mass 
%becomes a rather complicated function of the nuclear density. 

   In the zero nuclear density limit, 
%\footnote{Technically, in the spirit of 
%Refs.\cite{KAP86, NEL87,MAO99, MIS204}, this limit should be interpreted such
%that only one nucleon is left in the scene.}, 
the model is constructed such that the in-medium scalar meson contribution in 
Eq.~(\ref{eq:meanfields})  reduces to the meson mass term in the total $D$ meson Lagrangian,
viz. $m_D^2\bar D D$, as  the free $D$ Lagrangian should have only the kinetic
energy part to begin with.  This is consistent  with 
the corresponding limit in the light quark condensate: $<\bar q q>_{nuc} \to <0|\bar q q|0>$
in QCDSR: recall Eq.~(\ref{eq:condensate}). However, there is a touchy issue that 
needs clarification regarding the
 $D$ interaction with the mean vector meson field.  At zero nuclear density limit 
it is plausible that this contribution vanishes so only the T-W interaction remains.
But as will be touched upon later, the latter arises from  vector 
meson exchanges between $D$ and $N$ in the low
energy and low momentum transfer limit as inferred  from the Hidden Local
Symmetry picture of vector mesons \cite{BAN88}, or from the success, for
example,  of the $\rho$-$\omega$ model for the $KN$ interaction \cite{SIE88}
(see also \cite{BRO93}). So at finite nuclear density  there may well be some double 
counting in the vector meson exchange contribution.  In this respect we refer to an
interesting finding  in Ref.~\cite{MIS204}.  There a transport equation
simulation for {\it heavy-ion on heavy-ion} has been compared with available
data in the spectra of $K^+$ and $K^-$ produced in the reactions. The
model-data consistency has found to become troublesome  upon including the in-medium 
kaon-$\omega$ interaction. This might actually point to
an inadequacy of including the mean vector meson interactions for the case
of the $D$ meson as well. 

  In the end, the quantitative details of the NMFA prediction on the $D$ mass
shift vary depending on the details of models adopted, see \cite{MIS204}.
However, a global feature is characterized by 
moderate drop in the mass, obtained for the simple chiral Lagrangian 
of Eq.~(\ref{eq:KN}), viz.
about $70$ MeV for $D$ and $20$ MeV for $\bar D$ at normal nuclear
matter density. The average of those two values is a rough measure of the
scalar-isoscalar interaction. So just as from QCDSR, one sees a potentially 
important role played by this attractive force between the $D$ and
nucleon.

Another mean field approach which directly solves for the $D$ meson binding
in  nuclear medium is presented in \cite{TSU99, SIB99} within the
quark-meson coupling model \cite{GUI88}. Schematically, it is the "$\sigma$-$\omega$"
at the level of the $u$- $d$-quarks confined in nucleon and $D$-meson bags. 
More concretely, upon optimizing the mean scalar ($\sigma$) and vector ($\rho$ and $\omega$) 
meson fields by reproducing the static properties of nuclei (or nuclear matter), 
those meson fields are used to describe the in-medium $D$ interaction at the 
light quark level. 
Note that by construction there is no explicit T-W  interaction between   $D$
and $N$. The in-medium $D$($\bar D$)  mass is obtained by a
stationary condition upon varying the heavy meson bag radius. It is found that 
the magnitude of the vector and scalar  contributions to the
mass shift are comparable in magnitude. In particular, the  average
mass shift of the $D$ and $\bar D$ mesons due
to the scalar meson interaction is about $60$ MeV downward for
normal nuclear matter  density. 
Again, the importance of the attractive scalar-isoscalar  interaction
(represented here by the $\sigma$-meson exchange) is visible here.

Before reviewing the self-consistent coupled-channel approach (SCCM), we
simply summarize the common feature of the result  from QCDSR and NMFA
discussed so far: (i) the D-meson interactions are due both to static in-medium 
scalar and vector type interactions, and 
(ii) a large reduction in the $D$ meson mass 
to which a  scalar-isoscalar attraction appears to play an important role.

\subsection{SCCM}

The approaches discussed so far are $static$  in that the meson 
 nuclear interaction is introduced such that it does not disturb the  mean nuclear configuration. This is achieved when the meson
scattering is elastic and nearly forward by each nucleon (and by mean meson fields)
in nuclear  medium. Also  the meson-nucleon scattering should be
reasonably weak. In this respect we recall that in NMFA the meson self energy 
is obtained from the meson-nucleon interaction (or potential) 
rather than from its full iteration: the $T$-matrix.
Therefore, the methods would  become inappropriate when the two-body
meson-nucleon interaction  is  (i) strong and, in particular, dominated by
intermediate bound or resonant states, and/or (ii) strongly coupled to  other
meson-baryon channels.  
 The low energy ${\bar K}N$ interaction is a typical case that does not fit 
into the static mean-field description.  Its
coupling to other meson-baryon channels  such  as $\pi \Sigma$, $\eta \Lambda$,
... , etc. is strong.  Close to threshold, this interaction is dominated by the
nearby $\Lambda(1405)$ resonance,
which is now strongly believed to be a combination of ${\bar K}N$
and $\pi\Sigma$ $s$-wave molecules \cite{JID03,magas} embedded in the
continuum of lower threshold channels. Because of the apparent  similarity
between the  ${\bar K}N$  and $DN$ systems in their coupled channel nature as
well as  their association with the $\Lambda$ like  resonances:
$\Lambda(1405)$ and $\Lambda_c(2593)$, we think it  very useful to outline
the SCCM  used in the study of $\bar K$ in nuclear matter.  A good
part of its practical aspects is effective for  our  present 
 in-medium  $D$ study. See a  prototype of this approach in
\cite{KOC94}.

 Here again one needs two steps to achieve  the goal. But unlike NMFA the first step is for the two-body aspect,  then the  many-body aspect
enters later.  First, the free space  ${\bar K}N$ multi-channel
Bethe-Salpeter equation:
\begin{equation}
T=V+VGT \ ,
\label{eq:BSfree}
\end{equation}
is solved. Here, $T$ is the transition
operator matrix, $G$ a diagonal matrix  each element of which is  the  product of single particle  propagators for a  meson and a baryon, and the 
potential (driving term)  $V$ is a matrix whose elements are T-W type
meson-baryon interactions from the lowest order terms in the non-linear
chiral Lagrangian. In on-shell approximation the solution $T$  was shown to
successfully reproduce the $\Lambda(1405)$ resonance and other reaction
observables \cite{OSE98,LUT02}.  
%Note that the  diagonal
%$\bar KN$ \mbox{T-W} interaction in the $I=0$ channel alone is found sufficiently strong 
%to produce a {\it primitive} ``$\Lambda(1405)$'' bound state \cite{KOC94}, thus indicating
%the inadequacy of the  ``$\rho_N V$ approximation'' in NMFA to the low energy in-medium $\bar K$ self-energy.

Next, the same set of equations is solved in nuclear medium. The underlying 
assumption is  that the potential term $V$  is unaltered in nuclear medium, but
that all the medium effects enter through the intermediate meson-baryon
propagators \cite{LUT98, RAM00, LUT202, TOL02, ROT05, TRO06}. This in-medium
propagator  includes the effects from (i)  Pauli blocking,  (ii) binding of baryons by nuclear mean field, and (iii)  dressing (or self energy) of intermediate state mesons ($\bar K$,
$\pi$, or $\eta$) due to their interactions with surrounding nucleons. The
resultant quantity is denoted as $\tilde G$.  The in-medium equation now reads
\begin{equation}
\tilde T = V + V \tilde G \tilde T\ .
\label{eq:BSmedium}
\end{equation}
Because of the meson dressing, particularly the dressing of $\bar K$  which
creates  nested ${\bar K}N$ interactions, this in-medium equations must
be solved {\it self-consistently}. 
The thus obtained diagonal  amplitude, $\tilde T({\bar K}N \to {\bar
K}N)$, already demonstrates certain essential features of in-medium $\bar K$.
But a more suitable quantity to study is the kaon spectral function $S_{\bar
K(K)}(p_0, \vec p\, )$ which is proportional to the imaginary part of
the kaon self-energy
$\Pi_{\bar K(K)}(p_0, \vec p\, )$.  
In free space, we have a trivial on-mass
shell relation: $S_{{\bar K}(K)}(p_0, \vec p\, )=\delta(p_0-E(\vec p\,
))/(2p_0)$, with  $E(\vec p\, )= \sqrt{m_K^2+{\vec p\,}^2}$. In the nuclear
medium, this structure changes substantially, such that not only is the ${\bar
K}$ {\it meson mass pole}  shifted somewhat downward, but is broadened.
Moreover, there is an  additional structure due to the in-medium $\Lambda
(1405)$ resonance too \cite{LUT98, RAM00, LUT202, TOL02, ROT05, TRO06}. These aspects cannot be
obtained from the NMFA or QCDSR approaches discussed earlier. Notably, within SCCM there has  not been any  indication of a
possible  onset of kaon condensation even at higher densities. So one sees the
possible importance of medium effects taken care of self-consistently  
which have made
the difference.  On the other hand this difference might also be due, in part
or to a good extent, to the fact that so far the equations in SCCM   have been
driven only by the T-W vector interaction  without any additional ones such as
the attractive $\Sigma_{KN}$-term in the  diagonal ${\bar K}N$ channel present
in NMFA. Recently, there have been several works on the {\it improved} coupled
${\bar K}N$  equations in free space which incorporate the  {\it next to
leading order} interactions including the corresponding scalar-isoscalar (or
$\Sigma_{KN}$  term) contribution \cite{BOR05, OLL05, OLL06, BOR06}. Such
additional terms have certainly improved the fit  to available data by $\approx
20$\% thanks to several  additional  parameters related to these additional terms. With such a new type 
of interactions, one may wonder if the kaon mass
could reduce sufficiently in the nuclear medium to give rise to kaon condensation.
However, according to a recent work on in-medium $\bar K$ in isospin
asymmetric matter which incorporates this type of contribution  \cite{ROT05},
kaon condensation does not appear to set in below  eight times the nuclear matter saturation density. 
This is well beyond the limit of applicability of the model: certainly around this {\it critical} density 
new  degrees of freedom, both
hadronic and sub-hadronic, will have to be taken into account. 

 We now want to refer to a couple of exploratory works on the  $D$
meson within the same framework. First, a coupled-channels
calculation for the  $D$ mesons in cold nuclear matter has been 
done \cite{TOL04}, motivated by the similarity
between the $DN$ and ${\bar K}N$ systems once the $s$-quark in the
later is replaced by a $c$-quark. This is
further reinforced by  an apparent correspondence between the two  $I=0$,
$s$-wave resonances: $\Lambda(1405)$ and $\Lambda_c(2593)$, in the coupled
${\bar K}N$ and $DN$ channels, respectively. To make this analogy  more
concrete, free space amplitudes are obtained from a set of
separable coupled-channel potentials simulating the T-W type
interactions to
reproduce the  $I=0$ $\Lambda_c(2593)$ as a $DN$ $s$-wave {\it hadronic
molecular}  state of binding energy of $\approx 200$ MeV with a width of 
$\approx
3$ MeV, sitting very close to the $\pi \Sigma_c$ threshold.    In the $I=1$,
$s$-wave channel, the model appears to have generated a resonance at about
$2800$ MeV, but is less conspicuous than the fitted one in the $I=0$
channel.  Then the corresponding interaction is fed into the in-medium
equation. A notable
feature is the relative importance of the intermediate state pion dressing.
The final result has found a slight upward-shifted and  broadened $D$
meson $pole$ with a wiggle in the spectral function at normal nuclear matter
density. One of the peaks in the wiggle corresponds to the
shifted-broadened $D$ pole, while the other seems to originate from the resonant
structure in the in-medium $I=1$ $DN$ amplitude. 
Somewhat
surprising is the apparent absence of an anticipated peak due to the
$\Lambda_c(2593)$ resonance. The peak is visible when pion dressing is
ignored in the coupled channels problem.
This work has been recently extended to finite temperature in \cite{TOL06}.

Here a question remains as to  if the simple prescription
of $s \to c$ quark replacement be adequate to model the $DN$  coupled-channel
interaction.  By so doing all the two-body channels with strangeness, such as
$D_s\Lambda$, have been excluded. However, as we will show later,
 they have an important effect in the $DN$ coupled channel problem.
In addition, from the point of view of symmetries, one must recall  the  well-known fact:  while the light Goldstone bosons
such as $\pi$ and $K$ mesons are dictated by chiral symmetry,  the charmed
mesons such as $D$ are quite heavier and obey the {\it heavy-quark symmetry}:
just the extreme opposite to the former.  A blind $s \to c$ replacement breaks both
of those symmetries.  
%Hence,  the approach in Ref.\cite{TOL04} 
%should consequently be improved.

A different approach, which respects the proper symmetries,
has been attempted in Ref.~\cite{LK04}. There, charmed baryon
resonances are generated dynamically from the scattering of Goldstone bosons
off ground-state charmed baryons with $J^P=\frac{1}{2}^+$. The $C=1$, $S=I=0$
resonance found at 2650 MeV has been identified with the $\Lambda_c(2593)$ in spite
of the fact that
the width, due to the strong coupling to $\pi\Sigma_c$
states, is obtained as more than twenty times  the experimental value
of about 4 MeV. The trouble with this model is that couplings to 
$DN$ and  $D_s Y$ are completely absent.  We recall that 
in Ref.~\cite{TOL04} the former channel is essential 
in the formation of the $\Lambda_c(2593)$.

  A satisfactory improvement came in a recent work
\cite{HOF05} where the alleged  shortcomings  have been  overcome by
exploiting the universal vector meson coupling hypothesis to break the $SU(4)$
symmetry in a convenient and well-defined  manner. More precisely, this is done
by a $t$-channel exchange of vector mesons between pseudoscalar mesons and
baryons in such a way  to respect  chiral symmetry for the light meson sector
and the heavy quark symmetry for charmed mesons,  as well as to maintain the
interaction to be of the T-W vector type. 
%This allows for a natural
%communication between the two types of  meson-baryon sectors (the one
%with the charm quark
%belonging to the baryon and the other belonging to the meson) with the same
%total quantum number by a $t$-channel transfer of charm.
%According
%to \cite{HOF05}, 
The model generates  the $\Lambda_c(2593)$ resonance in the
$C=1$, $S=I=0$ $s$-wave channel, as well as an
almost degenerate $s$-wave resonance at $2620$ MeV in the
$C=1,\  I=1,\ S=0$  channel not found
experimentally so far. An application of this model to a preliminary study of
$D$ and $D_s$ mesons in nuclear matter may be found  in Ref. \cite{LUT05}. 
There  the $D$
meson  spectral function is found to  have a  peak, embedding the two resonances
mentioned above, and another structure signaling the $D$ meson pole
position, which is shifted upward by about $30$
MeV and has a finite width.   So, with respect to the $D$ pole mass shift, the two
SCCM results show the opposite tendency  to the one  found both
in QCDSR and NMFA, upon disregarding other attributes such as  finite widths,
etc. Then one may wonder from where this difference should originate: is it
attributable to the coupled-channel aspect, or to the absence of an extra
scalar-isoscalar attraction in the two SCCM models?

Having discussed salient aspects of various models to date, we have naturally
adopted SCCM in the present study of the $D$ meson in symmetric nuclear matter.
In particular, we have employed a modified version of the coupled channels
method developed by Hofmann and Lutz \cite{HOF05} after its critical analysis 
in the next section. Specifically, we have introduced a
cut-off regularization as well as an extra phenomenological scalar-isoscalar
 attraction in the diagonal $DN$ channel in a simplistic manner to study
its implication.

\section{Why have we adopted, but modified the Hofmann-Lutz model?}
\label{sec:review}

The original Hofmann-Lutz model \cite{HOF05} is  ambitious enough  to include
all the $J^P=\frac{1}{2}^-$ s-wave pseudoscalar-baryon interactions with
charm quantum number up to three
%: $C_{max}=+3$, 
in an attempt to
interpret/predict various baryon resonances as molecular states. Our present 
interest in this model is  only in the
sectors with  quantum numbers $C=1$, $S=0$, $I=0$ and $I=1$ which are
associated with the  $DN$ channel.  As for the $\bar DN$ sector with quantum
numbers $C=-1,  S=0$, QCDSR
\cite{HIG00, MOR02},  quark-meson coupling  model \cite{TSU99,SIB99}
as well as
simple quark model suggest that  the interaction is weak and quite likely repulsive.
See also Ref.~\cite{LUT05}. Hence we are not concerned with this sector in the present work.

 The Hofmann-Lutz  model connects two sets of characteristic pseudoscalar
meson-baryon sectors with one charmed quark belonging either to mesons, e.g.
$DN$, or to baryons, such as $\pi \Sigma_c$, etc. by means of the
universal vector
meson coupling hypothesis equipped with the KSFR condition \cite{KAW66}.
Its modern theoretical support is offered in the
hidden local symmetry picture of vector mesons, see  Ref.\cite{BAN88}. In this
respect the model is an important first step for improvement since heavy-quark
effective theory equipped with chiral symmetry, Chiral Heavy-Quark  Effective
Theory ($\chi$HQET), can only deal with  Goldstone bosons interacting  with
charmed baryons (or charmed mesons), see \cite{LK04} and 
\cite{WES92, BUR92, YAN92}. 
It cannot be applied to channels of our interest such as  $DN$.
While retaining the physical hadron masses, the Hofmann-Lutz  model 
uses $SU(4)$ 
symmetry to construct the effective interaction between 
pseudoscalar mesons in $16$-plet with baryons in $20$-plet
representations through a $t$-channel exchange of a
$16$-plet of vector mesons.
%The Hofmann-Lutz paper presents an extensive account of the model, to
%which the reader is referred for details. Here we just  outline its 
%feature useful for our objective. First, we note  that although $SU(4)$
%symmetry among  $u$-, $d$-, $s$-, $c$-quarks  is  badly broken regarding their current masses, it may still be possible to use it as a guide to
%construct the sought  effective meson-baryon interactions. Thus using this
%symmetry, pseudoscalar mesons in a $16$-plet and baryons in  $20$-plet
%representations  are made to interact  through a $t$-channel exchange of a
%$16$-plet of vector mesons. 
The universal vector  meson coupling
hypothesis provides the global interaction  strength among the above
$SU(4)$-multiplets. 
%The effective coupling  strengths in the sought  interactions
%are found in comparison with phenomenologically constructed effective
%Lagrangians in each relevant  $SU(3)$-{\it sliced} sector. Physical hadron
%masses are retained within this procedure. 
Then, aided by the {\it KSFR}
relation which is  consistent with  chiral symmetry at very low energy and
momentum transfer, the resultant lowest order meson-baryon interaction is
found  to take a near T-W form  in the $t=0$ limit. An interesting  and
important consequence of this picture is that, as compared with the ones which
have exploited  $SU(4)$ symmetry alone, the T-W
interactions resulting from an  exchange of a charmed meson are  reduced by an
extra factor $\sim (m_V/m^c_V)^2$, where the masses  here are the typical
(uncharmed) vector meson and a (singly) charmed meson, respectively. We shall 
discuss this aspect further below.

Now  we are in the position to obtain the interaction $V$ in our present work
based, in good part, on the Hofmann-Lutz model. The two sectors  of our interest
are all $s$-wave and have $J^P=\frac{1}{2}^-$.  We retain
the following channels: 
\begin{eqnarray*}
\pi \Sigma_c (2589), \ \mbox{\boldmath{$DN$}}(2810),\ 
\eta \Lambda_c(2835),
\ K \Xi_c (2960), \ K \Xi_c^\prime(3071), 
\ \mbox{\boldmath{$D_s \Lambda$}}(3085), 
\ \eta^\prime \Lambda_c(3245) \ ,
\end{eqnarray*}
for the $C=1,\ I=S=0$ sector, and
\begin{eqnarray*}
\pi \Lambda_c(2425),\ \pi \Sigma_c(2589), \ \mbox{\boldmath{$DN$}}(2810),
\ K \Xi_c(2960), \ \eta \Sigma_c(3005),\ K \Xi^\prime_c(3071),
\ \mbox{\boldmath{$D_s \Sigma$}}(3160), \hfill \hspace{2.5cm} \\
\eta^\prime \Sigma_c(3415) 
\phantom{
,\ \pi \Sigma_c(2589), \ \mbox{\boldmath{DN}}(2810),
\ K \Xi_c(2960), \ \eta \Sigma_c(3005),\ K \Xi^\prime_c(3071),
\ \mbox{\boldmath{$D_s \Sigma$}}(3160), \hfill \hspace{2.5cm}
} \ , 
\end{eqnarray*}
for the $C=1,\ I=1,\ S=0$ sector. Here, channels containing charmed pseudoscalar 
mesons are denoted in bold letters. The values between parentheses 
following each channel 
in the above expressions are the corresponding channel thresholds in MeV. 
The transition interaction (potential) for
$i\  \leftrightarrow \ j$ due to  $t$-channel exchanges of vector meson ``$X$'' reads (note a somewhat different notation from \cite{HOF05})
\begin{equation}
V_{ij}(q_i,q_j;\sqrt{s})=\sum_X g^2 C^X_{ij}\bar u(\vec p_i)
\gamma^{\mu}\left [g_{\mu \nu}-\frac{k_{\mu}k_{\nu}}{m_X^2}\right ]\frac{1}{t-m_X^2}(q_i+q_j)^{\nu}u(\vec p_j),
\label{eq:tchannel}
\end{equation}
where $g\ (\approx 6.6)$ is the universal vector meson coupling constant,
$C^X_{ij}$ is the product of relevant $SU(3)$ Clebsch-Gordan  coefficients 
associated with quantum numbers of the vertices
projected to the  $ij$ channels,  and   $p_i$, $q_i$, $p_j$, $q_j$ are 
the four momenta of the baryon and meson in channels $i$ and $j$, respectively. As
usual, $s=(p_i+q_i)^2=(p_f+q_f)^2$, and the momentum transfer is $k\equiv
q_i-q_j=p_j-p_i$, with $t\equiv k^2=k^{\mu}k_{\mu}$. 

The next step is to expand
$1/(t-m^2_X)$ in powers of $t/m^2_X$.  The second term in this expansion, viz.
$t/m^2_X$, tends to compensate to a good extent the term
$k_{\mu}k_{\nu}/m^2_X$ in the  numerator of the vector meson propagator, so to
a good approximation, the $t$ dependence is at most $O\left [(t/m^2_X)^2\right
]$.   Thus for our objective  it is consistent to disregard the terms
$O(1/m^2_X)$ altogether, and make the interaction to be of zero range. Note that 
the Hofmann-Lutz model has retained the
$O(1/m^2_X)$ term in the numerator.  The effect due to this additional
contribution will be examined later, particularly   for
charmed vector meson exchanges. Then we adopt the average mass $\bar m_V$ for
all the uncharmed nonet vector mesons as well as $\bar m_V^c$ for  the
anti-triplet charmed vector mesons. The difference caused by this
simplification is found
to be quite small.  By dropping the $p$-wave contribution from $\bar u(\vec
p_i)u(\vec p_j)$, we now have an $s$-wave, zero-range limit of the above
interaction in  the following T-W form upon adopting
the normalization convention ($\bar uu = 1$) used in \cite{JID03},
\begin{equation}
V^I_{ij}(\sqrt{s})= -\frac{\kappa C_{ij}}{4f^2}(2\sqrt{s}-M_i-M_j) 
\left(\frac{M_i+E_i}{2M_i}\right)^{1/2}\left(\frac{M_j+E_j}{2M_j}\right)^{1/2}
\ .
\label{eq:TWint}
\end{equation}
Here the left hand side is the $ij$ element of the (on-shell) interaction
matrix $V$ in  Eqs.~(\ref{eq:BSfree}) and  (\ref{eq:BSmedium}). On the right
hand side, $C_{ij}=\sum_X C^X_{ij}$,  $M_i$ and $M_j$ as well as $E_i$ and
$E_j$ are the masses and energies  of baryons in channels $i$ and $j$,
respectively. In addition,    $f$ is the pseudoscalar meson decay constant from
the KSFR relation: ${\bar m_V}^2/g^2=2f^2$. We have also introduced  $\kappa$,
a reduction factor  which is $unity$ for transitions $i \leftrightarrow j$,
driven by (uncharmed) vector meson exchanges ($\rho,\ \omega,\ \phi, \ K^*$)
etc.  but is equal to  $\kappa_c=(\bar m_V/\bar m_V^c)^2$ for charmed vector
meson exchanges such as $D^*$ and $D_s^*$.   The transition coefficients
$\tilde C_{ij}\equiv \kappa C_{ij}$, which are symmetric
with respect to the indices, are listed in Tables \ref{table:iso0} and
\ref{table:iso1}. The reader may  notice that the  thus obtained  T-W
interaction strengths {$C_{ij}$}  are simply the consequence of $SU(4)$
symmetry {\it modulo} sign convention and that the vector meson exchange
picture has provided a definite pattern of breaking the $SU(4)$ symmetric
interaction.

\begin{table}[hb]
    \centering
    \caption{The $\tilde{C}_{ij}$ coefficients for the $C=1$, $S=0$ meson-baryon interaction
    for isospin $I=0$}
   \begin{tabular}{l|ccccccc}
     & $\pi\Sigma_c$ & $DN$ & $\eta\Lambda_c$ & $K \Xi_c$ & $K \Xi_c^\prime$ & 
     $D_s \Lambda$ &
     $\eta^\prime\Lambda_c$ \\
\hline
      $\pi\Sigma_c$ & 4 &
      $\sqrt{\displaystyle\frac{3}{2}}\,\kappa_c$ &
      0 & 0& $\sqrt{3}$ & 0 & 0 \\
      $DN$ & & 3 &
      $-\displaystyle\frac{1}{\sqrt{2}}\,\kappa_c$&
      0 & 0 & $-\sqrt{3}$ &
      $-\kappa_c$
      \\
      $\eta\Lambda_c$ & & & 0 & $-\sqrt{3}$ & 0 &
      $-\sqrt{\displaystyle\frac{2}{3}}\,\kappa_c$&
      0 \\
      $K \Xi_c$ & & & & 2 & 0 &
      $-\displaystyle\frac{1}{\sqrt{2}}\,\kappa_c$&
      0 \\
      $K \Xi_c^\prime$ & & & & & 2 &
      $-\sqrt{\displaystyle\frac{3}{2}}\,\kappa_c$&
      0 \\
      $D_s \Lambda$ & & & & & & 1 & $\displaystyle\frac{1}{\sqrt{3}}
\kappa_c$ \\
      \\
      $\eta^\prime\Lambda_c$ & & & & & & & 0\\
    \end{tabular}
    \label{table:iso0}
\end{table}
\begin{table}[ht]
    \centering
    \caption{The $\tilde{C}_{ij}$ coefficients for the $C=1$, $S=0$ meson-baryon interaction
    for isospin $I=1$}
   \begin{tabular}{l|cccccccc}
     & $\pi\Lambda_c$ & $\pi\Sigma_c$ & $DN$ & $K \Xi_c$ & $\eta\Sigma_c$ & $K \Xi_c^\prime$ & $D_s \Sigma$ &
     $\eta^\prime\Sigma_c$ \\
\hline
 $\pi\Lambda_c$ & 0 & 0 & $-\sqrt{\displaystyle\frac{3}{2}}\,\kappa_c$
 & 1 & 0 & 0 & 0 & 0 \\
      $\pi\Sigma_c$ & & 2 & $1\,\kappa_c$ &
      0 & 0 & $\sqrt{2}$ & 0 & 0 \\
      $DN$ & & & 1 & 0 &
      $\displaystyle\frac{1}{\sqrt{6}}\,\kappa_c$&
      0 & $-1$ & $\displaystyle\frac{1}{\sqrt{3}}\,\kappa_c$
      \\
      $K \Xi_c$ & & & & 0 & 0 & 0 &
      $\sqrt{\displaystyle\frac{3}{2}}\,\kappa_c$&
      0 \\
      $\eta\Sigma_c$ & & & & & 0 & $-\sqrt{3}$ &
      $\sqrt{\displaystyle\frac{2}{3}}\,\kappa_c$&
      0 \\
      $K \Xi_c^\prime$ & & & & & & 0 &
      $-\displaystyle\frac{1}{\sqrt{2}}\,\kappa_c$&
      0 \\
      $D_s \Sigma$ & & & & & & & 1 & $-\displaystyle\frac{1}{\sqrt{3}}\,\kappa_c$ \\
     $\eta^\prime\Sigma_c$ & & & & & & & & 0\\
    \end{tabular}
    \label{table:iso1}
\end{table}

 Before testing and then using the resultant interaction, we want to check one
important aspect of the approximation  we have made to reach the T-W
form. Let us take Eq.~(\ref{eq:tchannel}) for  non-diagonal transition
interactions  by  charmed  meson exchanges, and as an extreme case consider the
one for $DN \leftrightarrow \pi \Lambda_c$. Here  the variable: $t=k^2$, in
$1/\left [t-(m_V^c)^2\right ]$  is far  from zero but could be as large as
$t \approx  +M_N^2$ where $M_N$ is the nucleon mass. Also, in such a transition,
the  $k_{\mu}k_{\nu}/(m_V^c)^2$ contribution in the numerator, which we have
disregarded, might significantly affect  the magnitude of the driving term: in
the channel under consideration it reduces the size of the driving term by more
than $50$\% in  the energy  range of our interest. However,  we have confirmed
numerically  that in the same energy range those two contributions  tend to mutually   
compensate such
that neglecting the two together makes a maximum deviation of $O\left [
(t^2/({\bar m}_V^c)^4\right ]\approx 25$\% as  compared with the original $t$-channel
charmed vector meson exchange interaction projected on to the $s$-wave. Near and 
above the $DN$ threshold it is only $10$\% or less.
Thus as stated earlier, our procedure has turned out not only to be   simpler
but  more consistent than what is adopted in \cite{HOF05}.
Combined with  the reduction  factor $\kappa_c$ already multiplied to this type
of transitions, our ordinary T-W form of the interaction is consistent  with
the  lowest order chiral symmetry as well as  heavy quark symmetry by use of
the extended KSFR relation. In actual calculations we simply set $\kappa_c
=1/4$ and the resulting amplitudes are found stable against a small variation
around this value.

\begin{figure}[t]
\begin{center}
\includegraphics[width=14cm]{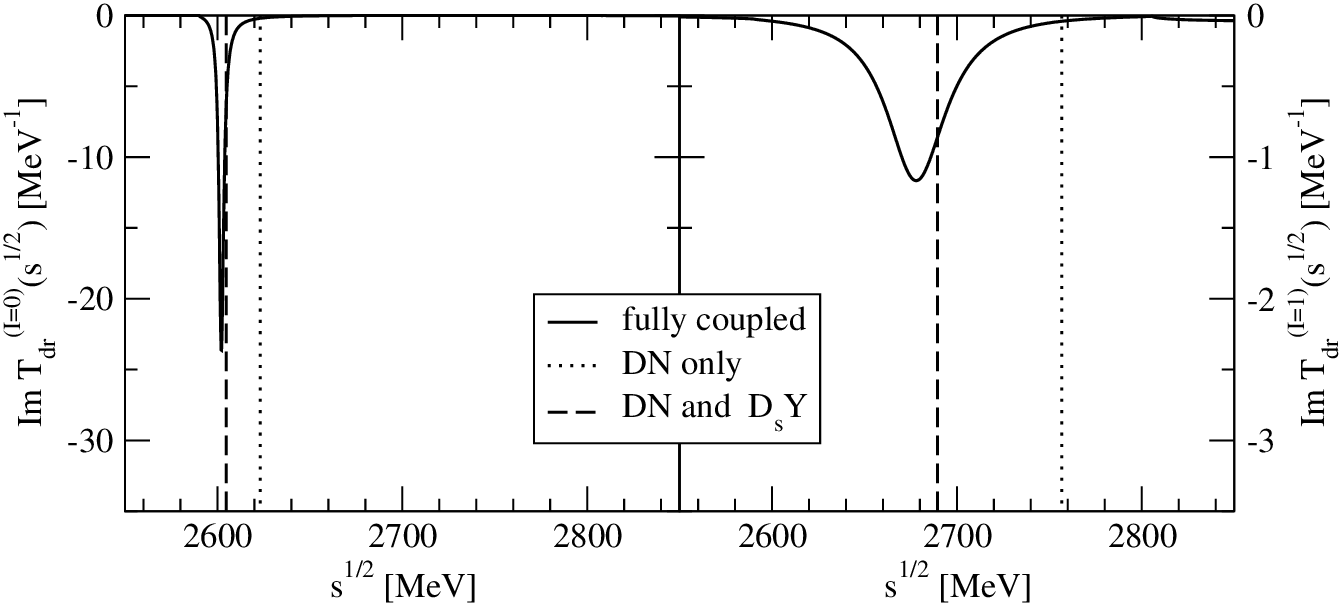}
\caption{Imaginary part of the $DN$ amplitude in the dimensional
regularization scheme as a function of $\sqrt{s}$ for $I=0$ (left
panel) and $I=1$ (right panel).}
\label{T_dr}
\end{center}
\end{figure}

The next step is to confirm the relevance of the interaction as obtained above.
We first check the resulting amplitudes in free space. Here an  on-shell ansatz (equivalent to the $N/D$ method)
has been employed which allows for reducing the coupled integral equations: Eq.~(\ref{eq:BSfree}), to a single matrix equation whose solution may be written as
\begin{equation}
T=(I-V\hat G)^{-1}V\,
\label{eq:BSfalgebra}
\end{equation}
where $\hat G$ is a diagonal matrix  whose  elements are now
four-momentum integrated propagators of the channels involved.   Momentum
integrations have been  regularized by a dimensional method as found in
\cite{ORB02,JID03}:
\begin{eqnarray}
G_{i}(\sqrt{s}) &=& i\, 2 M_i \int \frac{d^4 q}{(2 \pi)^4} \,
\frac{1}{(P-q)^2 - M_i^2 + i \varepsilon} \, \frac{1}{q^2 - m^2_i + i
\varepsilon}  \nonumber \\ &=& \frac{2 M_i}{16 \pi^2} \left\{ a_i(\mu) + \ln
\frac{M_i^2}{\mu^2} + \frac{m_i^2-M_i^2 + s}{2s} \ln \frac{m_i^2}{M_i^2} +
\right.  \\ & &  \phantom{\frac{2 M}{16 \pi^2}} +
\left. 
\frac{\bar{q}_i}{\sqrt{s}}
\ln\frac{M_i^2+m_i^2-s-2\bar{q}_i\sqrt{s}}
{M_i^2+m_i^2-s+2\bar{q}_i\sqrt{s}}
\right\}, \nonumber
\label{eq:gpropdr}
\end{eqnarray}
where $\bar{q}_i$ is the on-shell momentum, 
and $m_i$ and $M_i$ are the meson and baryon masses in
channel $i$. With
the regularization scale set to $\mu=1.0$ GeV and imposing
the subtraction points following \cite{HOF05}, we found that all
the corresponding subtraction constants  $a_i(\mu)$ stay close to {\it the
natural size}, viz. $\approx -2.0$ \cite{ORB02,JID03}. With this 
we calculate the
$C=1, \ S=0,\ I=0,1$ amplitudes, and find the
positions of the $\Lambda_c$ and $\Sigma_c$  resonances at  $\approx 2620$
MeV and $\approx 2680$ MeV, respectively, as in
\cite{HOF05} for the model they refer to as the $SU(4)$ {\it symmetric} case.
With a very small change in the  value of the subtraction constant
for the $DN$ channel in the $I=0$ sector, namely $a_{DN}(\mu): -1.92 \to -1.97$,
we were even able to adjust the position of the  resonance to
the empirical one for $\Lambda_c(2593)$. 
On the other hand,  a larger change
in the corresponding $a_{DN}(\mu)$ value appears to have been required  
to  shift the position of the $I=1$ resonance  down to $2620$ MeV as  predicted in
\cite{HOF05} for the model in which $SU(4)$
{\it symmetry} is broken upon shifting the value of the universal vector meson 
coupling
constant  by up to $\approx 20$ \%  in some channels. In the present work, we have  $\it not$
adopted  any such  modifications. Our results are shown by the solid
lines in Fig.~\ref{T_dr}. The meaning of the other two vertical lines will be
discussed in the next section. 

As for the widths of those resonances, our
values are far larger than those by  Hofmann and Lutz  for both isospin
sectors. The Hoffman-Lutz prediction for the $I=0$ resonance is no more than
$0.2$ MeV  while our estimate  is $\sim 3.0$ MeV,  closer to the experimental
value. A more dramatic difference is found in the $I=1$ sector:  the Hofmann-Lutz
prediction  is $\sim 3.3$ MeV or lower which is in sharp contrast to our 
large  estimated value of $\sim 35$ MeV.
  We have been able to trace the origin
of this difference to  the $k_{\mu}k_{\nu}/(m_V^c)^2$ term, retained in the
Hofmann-Lutz model. Since this term can reduce the charm exchange 
transition interaction by about a factor of two or more, the width
of the $DN$-type resonances decaying into $\pi \Lambda_c, \pi \Sigma_c$ states
are correspondingly smaller.
As discussed earlier, this term should not be
retained for consistency, hence our larger resonance widths should
be preferred. The prediction  for the position of the $I=1$ resonance will be
revisited in the next section  where we will introduce an explicit cut-off
regularization.

Upon confirming that  we have properly adopted the Hofmann-Lutz model with a
few simplifications, some of which have lead to improvements, we go on to 
include in-medium effects in the
amplitude. To achieve this we have followed Refs.\cite{LUT202} and
\cite{LUT05}, and solved
\begin{equation}
\tilde T=T+T(\tilde G-G)\tilde T\ ,
\label{eq:BSnoV}
\end{equation}
which results from combining
Eqs.~(\ref{eq:BSfree}) and  (\ref{eq:BSmedium}),
where we have included the Pauli
blocking effect as well as the $D$ meson dressing. Much of the  method of
solution has been taken from \cite{RAM00} and will be briefly described in
Sect.~\ref{sec:medium}.
Eq.~(\ref{eq:BSnoV}) must be solved self-consistently just like Eq.~(\ref{eq:BSmedium})
as the
dressed propagator $\tilde G$ contains the  solution $\tilde T$. So we have iterated the
equation by starting from the free space solution $T$. 
Although the input $T$ has been obtained by dimensional
regularization,  it is found that an explicit cut-off  must be introduced in
calculating  $(\tilde G-G)$ to extract a tempered in-medium solution. The
exception to this is when  only Pauli blocking is taken into account. Simply this
is because of the vanishing contribution to  $(\tilde G - G)$ from
momenta outside the nuclear Fermi sea. In
Refs.~\cite{LUT05} and \cite{LUT202} an explicit
cut-off value of $800$ MeV was used,  which appears reasonable
as an educated guess. However, it is important to check the stability of the
solution against the change in the cut-off value. So we have varied the upper 
limit of the momentum integration within the range accepted in effective hadron physics. 
As demonstrated in Fig.~2, the physics extracted in this way varies wildly as a function 
of the cut-off value, thus no reliable prediction of in-medium amplitudes appears possible. \\
In an attempt to overcome this trouble, we have gone back and
had a renewed look at the two original equations: 
Eq.~(\ref{eq:BSfree}), and Eq.~(\ref{eq:BSmedium}).
Regarding the free space one, viz. Eq.~(\ref{eq:BSfree}), one must remember that the resonance
positions and widths which it generates  depend inherently on how one
regularizes the divergent integral of the loop $G$. Employing a
dimensional regularization scheme, as in  \cite{HOF05} and in several other
works, has the advantage that the divergent part can be isolated analytically.
But for the in-medium equation, Eq.~(\ref{eq:BSmedium}), the divergent part cannot
be identified unambiguously as $\tilde{G}$ is available only numerically, so
one is forced to introduce an explicit cut-off. Then in the absence of any 
well-defined 
rule to relate the two regularization schemes, one would have no idea as to what 
cut-off value is relevant, nor to what extent the properties of the free space solution $T$ 
might have been carried over to the in-medium solution $\tilde{T}$. Furthermore, mathematically
the free and in-medium equations cannot be combined
into a single one, viz. Eq.~(\ref{eq:BSnoV}) when the support of the integration in the
two equations is not identical.
Hence, the method employed above to solve Eq.~(\ref{eq:BSnoV}) is an inconsistent one.
In view of this trouble,
we have discarded the dimensional method, but adopted the  {\it
conventional} cut-off method for the solution to the free-space equation as
well. This is described in the next section.

\begin{figure}[t]
\begin{center}
\includegraphics[width=14cm]{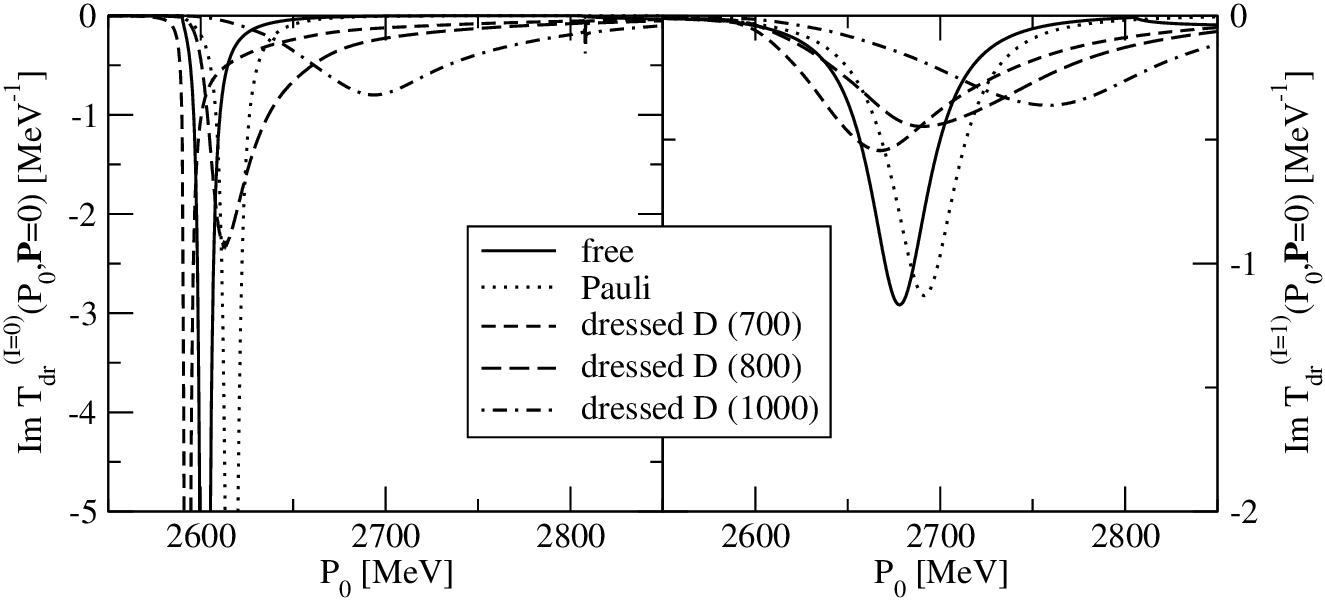}
\caption{Imaginary part of the in-medium $DN$ amplitude at total
momentum $\vec P=0$ as a function of total energy $P_0$ for
$I=0$ (left panel) and $I=1$ (right panel). The free amplitude
in the dimensional regularization scheme of Fig.~1
(solid lines) is
modified by medium corrections calculated according to Eq.~(\ref{eq:BSnoV}) with various
values of the cut-off: 700 MeV (short-dashed lines),
800 MeV (long-dashed lines) and 1000 MeV
(dot-dashed lines). The
dependence of the corrections on the cut-off is evident,
except when one only considers Pauli blocking (dotted line) where
the correction only affects loop momenta up to
$p_F=270$ MeV, well below any of the cut-off values
explored.}
\end{center}
\label{Tmed_dr}
\end{figure}

\section{Free space amplitudes in cut-off method}
\label{sec:free}

  From our discussion in the last section,  we have concluded that only the
direct cut-off regularization method is left to us as appropriate and practical
for our later study of the in-medium $D$ properties. So 
here we shall apply it identically to
both free and in-medium equations in a manner used  in \cite{RAM00} for the
study of $\bar K$ in nuclear matter.  In this section we construct a set of
free space amplitudes  in this scheme.  A  novel feature here is that 
we supplement
the T-W vector interaction discussed  in the previous section,
viz.  Eq.~(\ref{eq:TWint}), with  a scalar-isoscalar
attraction:  recall again our discussion on its possible importance in
Sect.~2.  Here we follow a simple and conventional  treatment of this term used
in the kaon condensate studies, and write it  as (see Ref.\cite{MIS104,MAR05} and
also the last term in Eq.~(\ref{eq:KN})):
\begin{equation}
 {\cal
L}_{\Sigma}\equiv \frac{\Sigma_{DN}}{f_D^2} \bar N N \bar D D\ .
\label{eq:Dsigma}
\end{equation}
In the above expression $f_D$ is the $D$ meson
weak decay constant, and $\Sigma_{DN}$ is the strength of this interaction.
%, but
%should not be confused with the familiar $\sigma$ terms appearing in the low
%energy $\pi N$ and $KN$ physics: recall our discussion in Sect.~2  about the
%Kaplan-Nelson  kaon condensates studies and the notation adopted therein.  
Note
that for simplicity we introduce this  only in the  diagonal $DN$ interaction:
presumably similar terms might claim their right in the diagonal $D_s Y,\
(Y=\Lambda, \Sigma)$ interactions since these channels couple strongly to $DN$
as understood from Tables \ref{table:iso0} and \ref{table:iso1}.  However, we
would like to look for possible  effects from such scalar-isoscalar attractions
in a semi-quantitative manner, thus prefer to contain the number of
parameters.
%In fact, combined effects from those coupled channels (including
%charmed mesons only) may likely be reasonably represented by the one in the
%$DN$ channel alone for our present purpose.
The $s$-wave projection of this
interaction is simply equal to
\begin{equation}
V_{\Sigma}(\sqrt{s})=-\frac{\Sigma_{DN}}{f_D^2}\left(\frac{M_N+E}{2M_N}\right)\ ,
\label{eq:sigma}
\end{equation}
for both $I=0,1$ $DN$ channels.
Concerning  $f_D$, its most recent extraction from the branching ratios of $D^+
\to \mu^+ \nu$ is about $223$ MeV \cite{ART05}. Various calculations and
measurements as cited in this reference do seem to agree within about $10$\%.
For simplicity we have adopted $f_D=200$ MeV.  As for the value of
$\Sigma_{DN}$, we simply  follow  what QCDSR \cite{HIG00} and NMFA of
\cite{MIS104} suggest and estimate it conservatively as $\Sigma_{DN} \approx
2000$ MeV.  We $also$ accommodate the case where no such attraction is added,
hence $\Sigma_{DN}=0$ then.

\begin{figure}[t]
\begin{center}
\includegraphics[width=14cm]{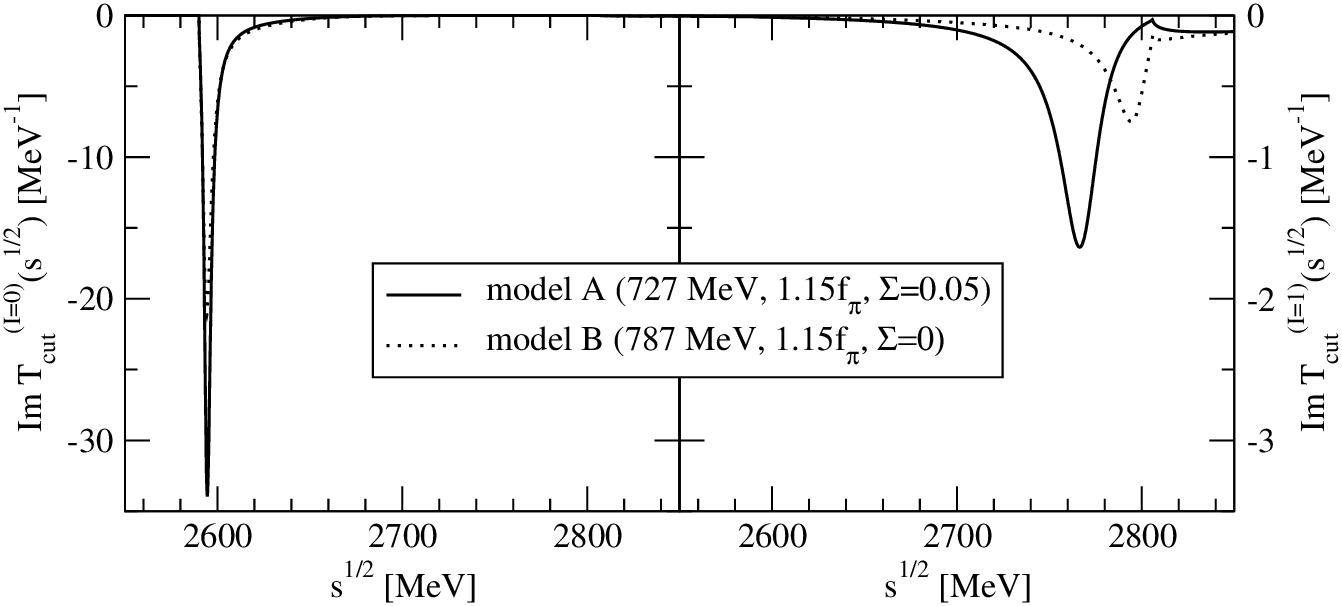}
\caption{Imaginary part of the $DN$ amplitude as a function of
$\sqrt{s}$ for $I=0$ (left panel) and $I=1$ (right panel), obtained by
using the cut-off scheme with model A (solid lines) and model B
(dotted lines)} \label{T_cut_2}
\end{center}
\end{figure}

With the above preparation, we solve the coupled-channel equations in free
space and reproduced the
$\Lambda_c(2593)$ resonance in the $I=0$ sector,
as seen on the left panel of Fig.~\ref{T_cut_2}, which shows the
imaginary part of the diagonal $I=0$ $DN$ amplitude for both models.
The parameters for those two cases are, 
model A: $f=1.15f_{\pi},\ \Sigma=\Sigma_{DN}/f_D^2=0.05$  MeV$^{-1}$, $\Lambda=727$ MeV,
and model B: $f=1.15f_{\pi},\ \Sigma=\Sigma_{DN}/f_D^2=0$ MeV$^{-1}$, $\Lambda=787$ \
MeV, where $\Lambda$ is the ultra-violet cut-off value for the integration in
the loop $G$ (the value of $f$ is loosely fixed in conformity with \cite{OSE98}).
%: somewhat easier in the case where $\Sigma_{DN}\ne 0$.
The width of the $I=0$ resonance
is found to be $\sim 4$ MeV for model A and $\sim
5$ MeV for  model B, respectively. 
Note that in the same isospin sector,
another resonance very close to $\Lambda_c(2593)$ but far wider has been
identified as in \cite{HOF05} which is due to the chiral excitation  in the
charmed hyperon channels \cite{LK04}, and is connected to the $DN$ channel by
a charmed vector meson exchange. So this is more easily seen in the  $\pi
\Sigma_c$ channel. The  same two sets of parameters are now used in the
$I=1$ coupled channels,  and we have found a somewhat different prediction as
seen in the right panel of Fig.~\ref{T_cut_2} for the
diagonal $I=1$ $DN$ amplitude. Although a resonance is generated in both models, 
the one with an extra  attractive $\Sigma_{DN}$ interaction (model A)
pulls the resonance lower  to about $2770$ MeV  (with a width $\sim 25$ MeV),
while the one without it (model B) keeps the resonance position at $\approx
2795$ MeV, not very far from the $DN$
threshold: $\approx 2810$ MeV, together with a width of $\sim 20$ MeV. 
Tentatively we call this $I=1$ resonance as $\Sigma_c(2770)$.

\begin{figure}[t]
\begin{center}
\includegraphics[width=14cm]{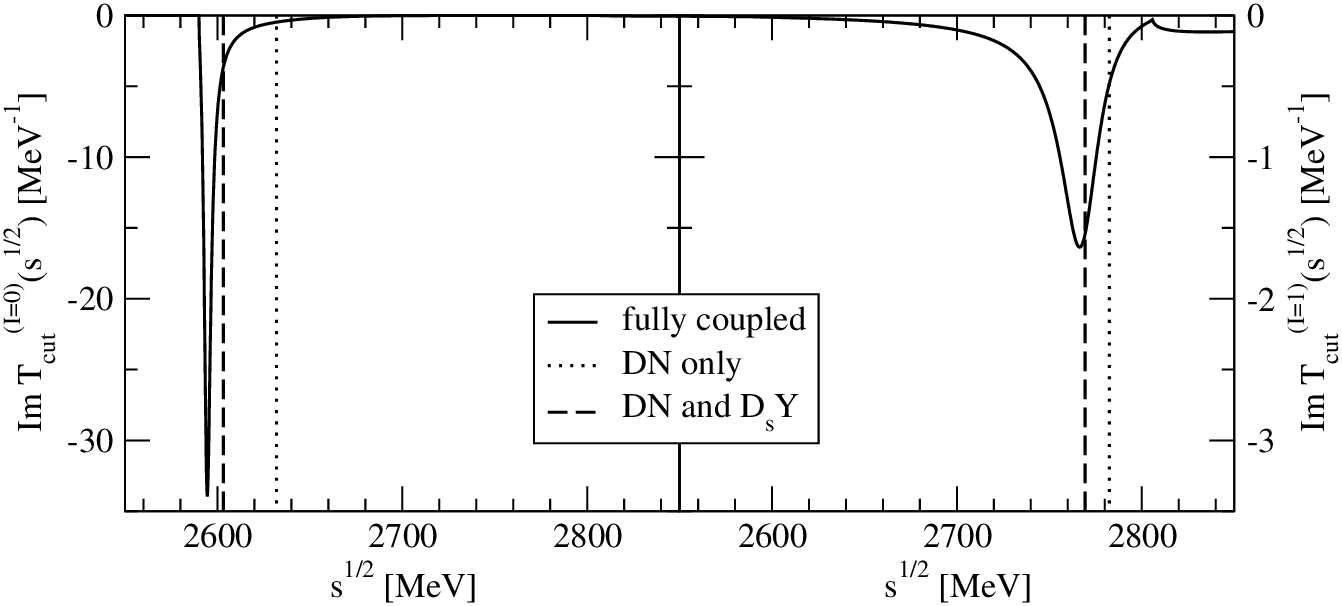}
\caption{Imaginary part of the $DN$ amplitude as a function of
$\sqrt{s}$ for $I=0$ (left panel) and $I=1$ (right panel) obtained by
using the cut-off scheme (model A) } \label{T_cut_1}
\end{center}
\end{figure}

To understand how the difference in the two isospin sectors comes about, it
should be useful to have an  {\it anatomical study} of the  underlying
mechanism for   resonance formation. The following argument can be made
qualitatively, based upon Tables \ref{table:iso0} and  \ref{table:iso1}, and
can then be
substantiated quantitatively by actual calculations. First, we see that the
T-W diagonal $DN$ interaction is  attractive in both isospin channels,
hence when channel couplings
are turned  off, the $s$-wave $DN$ state may form a bound state in both $I=0$
and $I=1$ sectors, as seen by the dotted lines in Fig.~\ref{T_cut_1},
corresponding to model A, although the same behavior is found for 
model B, viz. with and without the additional attraction by $\Sigma_{DN}$. The
binding is, of course,  deeper for $I=0$ because the T-W interaction
is three times stronger. Next, we introduce channel  couplings
$DN$-$D_s Y, \ (Y= \Lambda, \ \Sigma)$. Note that
this type of coupling is absent in
\cite{TOL04} where particles with strangeness  are excluded. In both isospin
channels the corresponding strengths are comparable to the $DN$ diagonal
interaction. Because the  thresholds for these $D_s$ involved channels are
higher than those with $D$, the channel coupling brings additional attraction
to the  $DN$ binding, an effect that can be easily justified from second order
perturbation arguments. Thirdly,  the remaining couplings with the $DN$
channel are from those connected by a charm transfer, hence suppressed by
$\kappa_c \approx 1/4$, so the additional shift in the bound state pole
positions is smaller. The only apparent
effect from coupling to $\kappa_c$-suppressed channels on the diagonal  $DN$
amplitudes is the finite width due to those channels with lower thresholds, hence
transforming the bound states to resonances.

By comparing the results of our explicit cut-off models as in
Figs.~\ref{T_cut_2} and \ref{T_cut_1} with the ones
from the dimensional  scheme
shown in  Fig.~\ref{T_dr}, one sees rather large differences in the $I=1$
sector. In the latter the downshift in the $DN$ {\it bound state} pole
position due to  coupling to the $D_sY$ channel is found larger than in
the $I=0$ sector, just the opposite to what one finds in  the explicit cut-off
scheme. In addition, the dimensional scheme finds the $I=1$ resonance
position at $2680$ MeV,  quite lower than the one from the cut-off method.
To understand the possible origin of the differences, we look at the
choice of the
subtraction point in  the dimensional approach as discussed in \cite{HOF05},
which was taken in each sector (with a set of definite
quantum numbers) at $\mu =\sqrt{m^2+M^2}$ where $m$ and $M$ are the meson and
baryon masses in the channel in which $m+M$ is minimum.
In the present $C=1$, $S=0$ case, $m=m_\pi$ for both isospin sectors while 
$M=M_{\Sigma_c}$
for $I=0$ and $M=M_{\Lambda_c}$  for $I=1$. This definition is somewhat
different from the one used in \cite{LUT04,GAR04}: presumably, the choice in
\cite{HOF05} would  make more sense numerically when both $m$ and $M$ are
relatively  large and/or comparable in size, but is not an absolute measure.
In due consideration of such a difference,  there does not seem to be any
reason to insist upon choosing different subtraction points  for the two
present isospin sectors where the difference in the value of the subtraction
constants is practically the pion mass,  $M_{\Sigma_c}-M_{\Lambda_c} \sim
m_{\pi}$. In addition, as stated in our anatomy study above, the $I=1$ $bound$ or
resonant state should  come visibly higher than  the one for the $I=0$ sector.
So it should make sense to adopt the subtraction point for $I=0$ also in the
$I=1$ sector {\it as  viewed from} the $DN$ channel.  In addition, such a minor
change will not disturb the approximate crossing symmetry as promoted in
\cite{HOF05}. In fact, when we made this increase in $\mu$: $\mu_{I=1} \to
\mu_{I=0}$, then the $I=1$ channel resonance
position  goes up  from $2680$ Mev to  $\sim 2750$ MeV, the new value being
closer to the one produced in our cut-off  scheme.
%So a small increase in the subtraction
%constant in the  Hofmann-Lutz dimensional scheme appears to correspond to a
%stronger cut-off. See Ref.\cite{GAR05} for a similar observation in a somewhat %different context.
The sensitivity of $I=1$ resonance pole position
to a relatively  small shift in the subtraction point had been least anticipated,
and thus is a little surprise.

Before finishing this section devoted to our study of the two-body input for
the in-medium calculation, we note that the Belle Collaboration has recently
measured in this energy range an isotriplet of excited charmed baryons decaying
into $\Lambda_c^+\pi^-$, $\Lambda_c^+\pi^0$ and $\Lambda_c^+\pi^+$
\cite{belle05}.  It is interpreted as a new charmed baryon, the
$\Sigma_c(2800)$, having a width of around 60 MeV, measured with more than
50 \% error.  This baryon has been tentatively identified with a $D$-wave resonance,
to conform to quark model predictions \cite{Copley79}, although the expected
width $\Gamma \sim 15$ MeV \cite{Pirjol97} is smaller than the
observed one. Actually, the fits performed in \cite{belle05} were not too
sensitive to varying the signal parametrization using $S$-wave or $P$-wave
Breit-Wigner functions, hence this resonance could still qualify as an $S$-wave
meson-baryon molecule of the type  found in the present  work around the same
energy and having a width of $\sim 40$ MeV, which is compatible with the experimental  one.

%From this exercise we assess that the predicted position of the $I=1$ channel
%resonance  could well be about $2680$ MeV and up  even within the scheme in
%Ref.\cite{HOF05} including the $SU(4)$ breaking in the couplings.  This
%estimate  appears reasonable, in particular,  from the coupling strengths of
%the T-W interaction in  Tables \ref{table:iso0} and \ref{table:iso1} along with
%our anatomical study  above.  Hence it looks rather unlikely to expect the
%resonance position in $I=1$ to become close  to the one for $I=0$. We note that
%within the dimensional scheme, the sensitivity of $I=1$ resonance pole position
%to a relatively  small shift in the subtraction point had been least anticipated,
%and thus is a little surprise.  This concludes our study in the two-body input
%for the in-medium calculation.

\section{$DN$ coupled channels equation in nuclear matter}
\label{sec:medium}

The first obvious medium effect to be included in the $DN$ coupled
channels equations is the Pauli
blocking on the intermediate nucleon states. This is a
particularly important one in the vicinity of a dynamically
generated resonance, as was explicitly shown in
Ref.~\cite{KOC94} for the $\Lambda(1405)$ in the context of ${\bar
K}N$ scattering. Here, intermediate nucleons need more energy to
access states that are not occupied, so the resonance is generated at
higher energies, moving from below to above the
${\bar K}N$ threshold. This induces, in turn, strong changes in the
${\bar K}N$ amplitude near the threshold. Namely, the threshold behavior of
the amplitude  changes from repulsive in
free space to attractive in the medium upon including Pauli corrections.
When this attraction felt by the ${\bar K}$ is now
fed into the solution of the in-medium amplitude,
the resonance moves back to lower energies. This thus  necessitates a
self-consistent calculation, which was done in Ref.~\cite{LUT98},
where the resonance was found practically at the
same location as in free space. This behavior was confirmed in
Ref.~\cite{RAM00} which incorporated also the self-energy of the pions
present in
the coupled channels problem along with the baryon binding.

Based upon the above observation on the in-medium ${\bar K}$ behavior,
we shall also consider those medium effects
in our present study on the properties of the $D$ meson following
the approach presented in Ref.~\cite{RAM00}.
%, which has been recently
%improved in Ref.~\cite{TRO06} with the inclusion of the $p-$wave
%meson-baryon interaction. It has also been explicitly shown there
%that, for a $s-$wave interaction as the one assumed in the present work
%[Eq.~(\ref{eq:TWint}], the on-shell factorization of the in-medium
%amplitudes can be undertaken as in free space.
For this purpose, we have only
 to incorporate them in the meson and
baryon propagators of the loop function $G$, which is then
denoted as $\tilde{G}$. Note that here we have not included
baryon binding energies altogether due to our lack of
knowledge of the charmed baryon mean-field potentials.

The effects of Pauli blocking are simply included by replacing the
free nucleon propagator by the in-medium one,
\begin{equation}
G_N(p_0,\vec{p}\,) =
\frac{1-n(\vec{p}\,)}{p_0-E_N(\vec{p}\,)+i\varepsilon} +
\frac{n(\vec{p}\,)}{p_0-E_N(\vec{p}\,)-i\varepsilon} \ ,
\label{eq:nuc}
\end{equation}
where $n(\vec{p}\,)$ is the nucleon occupation
with value 1 (0) for nucleons below (above) the Fermi momentum
and $E_N(\vec{p}\,)$ is the
nucleon energy. For the $D$ (and $\pi$) mesons we incorporate the
corresponding self-energy (dressing) in the propagator
\begin{equation}
D_D(q_0,\vec{q}\,) = \frac{1}{q_0^2-\vec{q}\,^2 - m_D^2 -
\Pi_D(q_0,\vec{q},\rho)} \ ,
\end{equation}
which is done, in practice, through the corresponding Lehmann
representation:
\begin{equation}
D_D(q_0,\vec{q}\,) = \int_0^\infty d\omega
\frac{S_D(\omega,\vec{q}\,)}{q_0-\omega+i\varepsilon} -
\int_0^\infty d\omega \frac{S_{\bar
D}(\omega,\vec{q}\,)}{q_0+\omega -i\varepsilon}
 \ ,
\end{equation}
where $S_{D({\bar D})}(q_0,\vec{q}\,)$ is the spectral function
of the $D$ ($\bar{D}$) meson.
In free space it is simply
\begin{equation}
S_{D({\bar D})}(q_0,\vec{q}\,) =\Theta (q_0)\delta(q_0^2-{\vec q\,}^{2}-m_D^2) 
\ ,
\end{equation}
where $\Theta (q_0)$ is the Heaviside step function. In the nuclear medium the
spectral function becomes
\begin{equation}
S_D(q_0,{\vec q}\,)= -\frac{1}{\pi} {\rm Im}\, D_D(q_0,{\vec q}\,)
= -\frac{1}{\pi}\frac{{\rm Im}\, \Pi_D(q_0,\vec{q}\,)}{\mid
q_0^2-\vec{q}\,^2-m_D^2- \Pi_D(q_0,\vec{q}\,) \mid^2 } \ .
\label{eq:spec}
\end{equation}
With these medium modifications the propagator loop function $\tilde{G}$
reads:
\begin{eqnarray}
\tilde{G}_{D N}(P_0,\vec{P}\,)= &&\int_{\mid {\vec q}\, \mid <
q_{\rm max}^{\rm lab}} \frac{d^3 q}{(2 \pi)^3}
\frac{M_N}{E_N (\vec{P}-\vec{q}\,)} \times  \nonumber \\
&&\left[ \int_0^\infty d\omega
 S_{D}(\omega,{\vec q}\,)
\frac{1-n(\vec{P}-\vec{q}\,)}{P_0- \omega - E_N
(\vec{P}-\vec{q}\,)
+ i \varepsilon} \right. \nonumber \\
&&+ \left. \int_0^\infty d\omega
 S_{\bar D}(\omega,{\vec q}\,)
\frac{n(\vec{P}-\vec{q}\,)} {P_0+ \omega -
E_N(\vec{P}-\vec{q}\,)-i \varepsilon } \right] \ ,
\label{eq:gmed}
\end{eqnarray}
for $DN$ states and
\begin{eqnarray}
\tilde{G}_{\pi Y_c}(P_0,\vec{P}\,)&= & \int_{\mid {\vec q}\, \mid < q_{\rm
max}^{\rm lab}} \frac{d^3 q}{(2 \pi)^3} \frac{M_{Y_c}}{E_{Y_c}
(\vec{P}-\vec{q}\,)} \int_0^\infty d\omega
 S_\pi(\omega,{\vec q}\,) \nonumber
\\
& \times & \frac{1}{P_0- \omega - E_{Y_c} (\vec{P}-\vec{q}\,) + i
\varepsilon}  \ , \label{eq:gmedpion}
\end{eqnarray}
for $\pi \Lambda_c$ or $\pi \Sigma_c$ states, where
$P=(P_0,\vec{P}\,)$ is the total two-particle four momentum and ${\vec q}$ is
the meson momentum in the nuclear matter rest frame. 

For $\eta
(\eta^\prime) Y_c$, $K \Xi_c (\Xi_c^\prime)$ and $D_s Y$ states, the corresponding
meson lines are not dressed by  self-energy insertions, viz. we use the
loop integral for free space. The reasons are: (i) the coefficients coupling the 
$\eta(\eta^\prime) Y_c$ channels with the $DN$ channel are small, 
as can be seen in Tables~\ref{table:iso0} and
\ref{table:iso1},  and (ii) containing an ${\bar s}$-quark, the $K$ couples
weakly to nucleons and its spectral function may be approximated by 
the free space one, viz. by a delta function. Note that this last prescription applies also to
$S_{\bar D}(\omega, \vec q \ )$ in Eq.~(\ref{eq:gmed}).  As for the spectral
function of the $D_s^+$ meson appearing in the in-medium $D_sY$ channels, it has been
shown \cite{LUT05} that,
in addition to the quasi-particle peak, it presents a lower energy
mode associated with an exotic resonance predicted around 75 MeV
below the $D_s^+ N$ threshold \cite{HOF05}. Therefore, with large coupling
coefficients for transitions $DN \leftrightarrow D_s Y$, as seen in Tables \ref{table:iso0}
and \ref{table:iso1}, one may eventually have to
solve an extended in-medium self-consistent coupled channels problem 
combining the
$C=1, S=0$ ($DN$) and  $C=1, S=1$ ($D_s N$) sectors. This arduous task will be relegated to a future work.

Now the in-medium  amplitudes $\tilde{T}$ are obtained
by directly solving the coupled-channel Eq.~(\ref{eq:BSmedium}) with the
medium modified loop function $\tilde{G}$ discussed above, or
from the equivalent Eq.~(\ref{eq:BSnoV}) where, formally, the medium
correction appears as the second term on the right hand side.

The in-medium $D$ self energy is finally obtained
by integrating  $\tilde{T}_{D N}$ over
the nucleon Fermi sea as
\begin{eqnarray}
\Pi_{D}(q_0,{\vec q}\,)&=&2\int \frac{d^3p}{(2\pi)^3}\,
n(\vec{p}\,) \,  [\tilde{T}_{D n}(P_0,\vec{P}\,) + \tilde{T}_{D
p}(P_0,\vec{P}\,)] \nonumber \\
&=& \int \frac{d^3p}{(2\pi)^3}\, n(\vec{p}\,) \,
[\tilde{T}^{(I=0)}(P_0,\vec{P}\,) +
3\tilde{T}^{(I=1)}(P_0,\vec{P}\,)]\ , \label{eq:selfd}
\end{eqnarray}
where $P_0=q_0+E(\vec{p}\,)$ and $\vec{P}=\vec{q}+\vec{p}$ are the
total energy and momentum of the $DN$ pair in the nuclear matter
rest frame and the values ($q_0$,$\vec{q}\,$) stand  for the
energy and momentum of the $D$ meson also in this frame. Recall that
$\Pi_D(q_0, \vec q\ )$ must be determined self-consistently since it
is obtained from the in-medium amplitude $\tilde T_{DN}$ which contains 
the $DN$ loop function
$\tilde G_{DN}$, and this last quantity itself is a function of
$\Pi_D(q_0, \vec q\,)$.

\section{Results and Discussion}
\label{sec:results}

Before starting our discussion on the various medium effects, let
us first explore what is the mass shift of the $D$ meson
in isospin symmetric nuclear matter for various simple approximations.
We define the effective mass as $m^*_D=\omega_{qp}(\vec{q}=0)$, where
the quasi-particle energy, $\omega_{qp}$, is the solution to the equation
\begin{equation}
\omega_{qp}(\vec{q}\,)^2=\vec{q}\,^2 + m_D^2 + {\rm
Re}\,\Pi_D(\omega_{qp}(\vec{q}\,),\vec{q}\,) \ . \label{eq:qpe}
\end{equation}
The mass shift can then be approximated by the $D$-meson optical
potential at zero momentum:
\begin{equation}
m_D^{*}-m_D \simeq U_D(0) \equiv \frac{{\rm
Re}\,\Pi_D(m^*_D,\vec{q}=0)}{2 m_D } \ .
\end{equation}

First, when one adopts a
mean field type of approximation and replaces $\tilde{T}_{DN}$ by 
the diagonal $V_{DN}$ in
Eq.~(\ref{eq:selfd}), the
mass shift is found as $\sim -60$ MeV for model A that
considers $\Sigma_{DN} \neq 0$ or $-43$ MeV for model B, viz. in the absence of
the $\Sigma_{DN}$ term. This amount of attraction is comparable to what have been
obtained in QCDSR and NMFA models as 
discussed in Sect.~2. \cite{HIG00, MOR02, TSU99, MIS104}.
This may be regarded as the calibration of our diagonal $DN$ interaction with
respect to those earlier works. 

Next we go one step to replace $V_{DN}$ by the free-space $T_{DN}$.
Due to the presence of resonances generated, the latter quantity is obviously distinct
 from the former. So we expect a better description than the first one by using $T_{DN}$ in
Eq.~(\ref{eq:selfd}), a procedure that
is referred to as the $T\rho_N$ approximation to the $D$ meson
self-energy. With this the $D$ meson feels a repulsion of
$\sim 25$ MeV, rather than attraction, indicating the importance of a
non-perturbative treatment of the problem once resonances are
present. As will be discussed later, in the present model this drastic change
is caused mainly by the isospin one $\Sigma_c(2770)$ resonance which lies below but close to
the $DN$ threshold.

Then we go further to consider the medium effects, Pauli blocking and the self-consistent
inclusion of the $D$-meson self-energy in the coupled channels
equations.  When  Pauli blocking alone is
considered, the amount of repulsion increases up to 40 MeV,
whereas when  meson dressings are incorporated in addition, the
repulsive mass shift goes down to 5 MeV. Note, however, that the
actual shift of the $D$-meson mass is eventually determined by the
position of the quasi-particle peak in the spectral function,
which has a complex structure determined both by the real and the
imaginary parts of the self-energy. We will come back to this
point towards the end of this section.

\begin{figure}[t]
\begin{center}
\includegraphics[width=14cm]{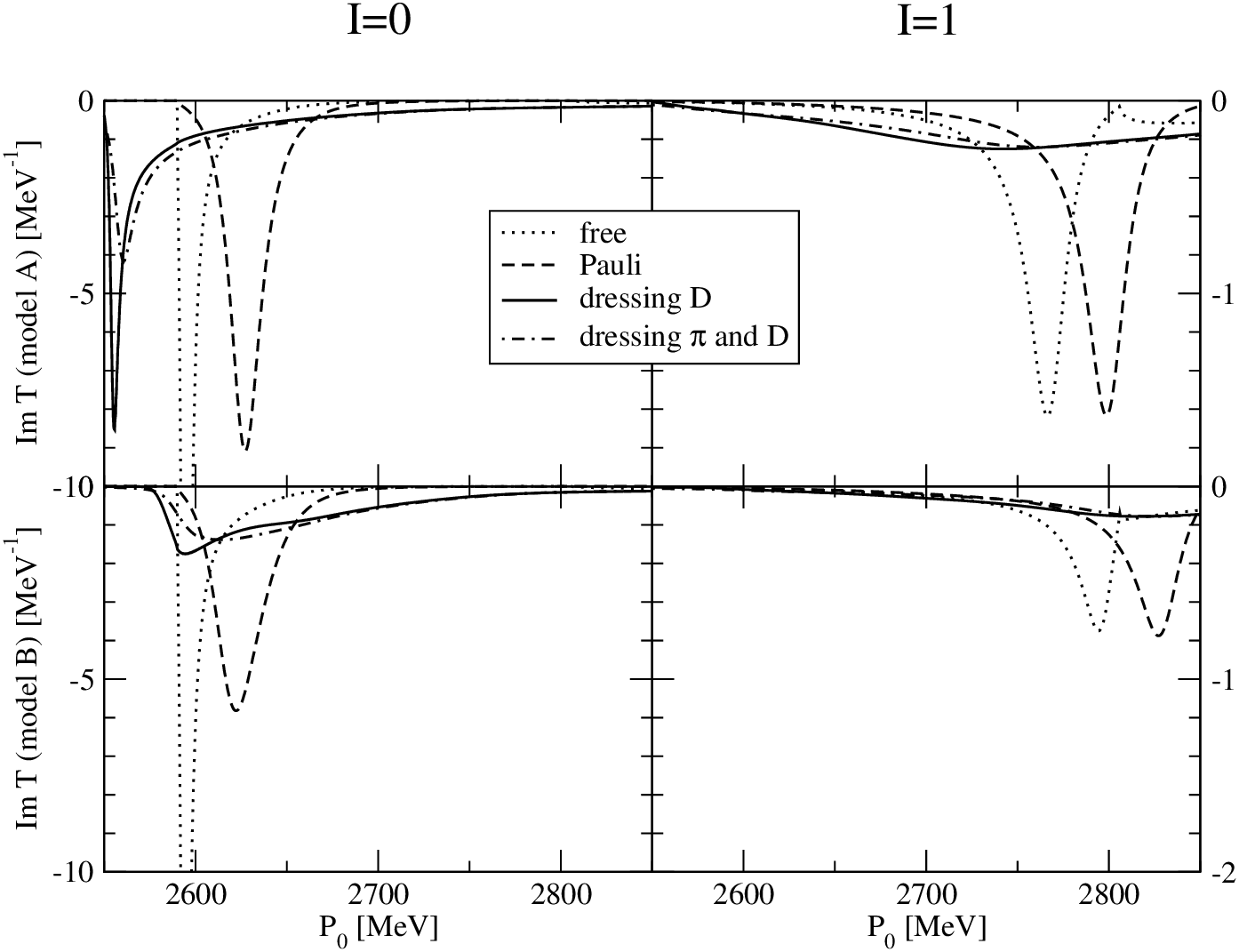}
\caption{Imaginary part of the in-medium $DN$ amplitude for total
momentum $P=0$  as a function of the total energy $P_0$ for $I=0$
(left panels) and $I=1$ (right panels), obtained in the cut-off
scheme using model A (upper panels) and model B (lower panels). }
\label{Tmed_cut_1}
\end{center}
\end{figure}

Being convinced of the importance of a proper treatment of in-medium
effects, we now discuss more explicitly the changes induced 
in the $DN$  amplitude and, as a result, on the $D$ meson self-energy 
and its spectral density. In
Fig.~\ref{Tmed_cut_1} we display the imaginary part of the $I=0$
(left panels) and $I=1$ (right panels) $DN$ amplitude at normal nuclear matter
density, $\rho_0=0.17$ fm$^{-3}$, as a function of the
total energy $P_0$ and total momentum ${\vec P}=0$. The results of
model A and B are presented in the two upper and two lower panels,
respectively. With Pauli blocking (dashed lines),
the resonances are produced at higher energies than in the free
amplitudes (dotted lines), in exact analogy to the behavior of the
${\bar K}N$ in-medium amplitudes described at the beginning of
this section. When one adds the $D$ meson dressing, the
$I=0$ and $I=1$ resonances move down (solid lines), even below
their corresponding free space location. This effect is especially
pronounced for model A that contains a non-vanishing
$\Sigma_{DN}$ term. In particular, the $\Lambda_c(2593)$ appears
about 50 MeV lower in energy than in free space. The reason for this additional attraction
when the $D$ self-energy is included self-consistently is that the $DN$
amplitudes develop strengths at  much lower energies than
their free space thresholds, starting actually at the threshold for
the $\pi\Lambda_c$ states, 2422 MeV. This enhances the phase-space for
intermediate states, inducing effectively a strong attraction in
the coupled channels equations. The result is the lowering
of resonance positions below their free space counterparts. These
in-medium resonances will be denoted from now on as $\tilde\Lambda_c(2593)$
and $\tilde\Sigma_c(2770)$. In particular, the
mass of the $\tilde\Lambda_c(2593)$ is lowered by about 50 MeV in nuclear matter at
normal density. 
Note that the width of the in-medium $\tilde\Lambda_c(2593)$ is not 
zero even if
it now appears below the free space $I=0$ coupled-channel threshold,
$\pi\Sigma_c$. This is due
to the process $\tilde\Lambda_c(2593) N \to \pi N \Lambda_c$, which
opens up as soon as the in-medium $D$-meson self-energy is incorporated. 
A similar argument holds for
explaining the much larger width of the $I=1$ resonance which
can also decay through
nucleon absorption processes of the type $\tilde\Sigma_c(2770) N \to
\pi N \Lambda_c, \pi N \Sigma_c$. When the pions are also dressed
(dot-dashed lines in the same figure), the tendency does not change
much even if
two-nucleon absorption channels, $\tilde\Lambda_c(2593) N N \to
\Lambda_c N N, \Sigma_c N N$, are now possible through the
coupling of the pion to particle-hole configurations. This is in
contrast to what is observed for ${\bar K}N$ dynamics \cite{RAM00}
and also to what is found in Ref.~\cite{TOL04} for in-medium
$D$-mesons. The reason is that, in the present model, the interaction
$DN \to \pi Y_c$ is reduced by the factor
$\kappa_c \simeq 1/4$ originating from the $t$-channel exchange of a charmed
vector meson.

The results for models A and B are qualitatively similar. The only
noticeable difference is that, due to the absence of
attractive scalar-isoscalar $\Sigma_{DN}$ term, model B produces in-medium
resonances at higher energies. Correspondingly, their widths are larger due
to the increased decaying phase space.

\begin{figure}[t]
\begin{center}
\includegraphics[width=14cm]{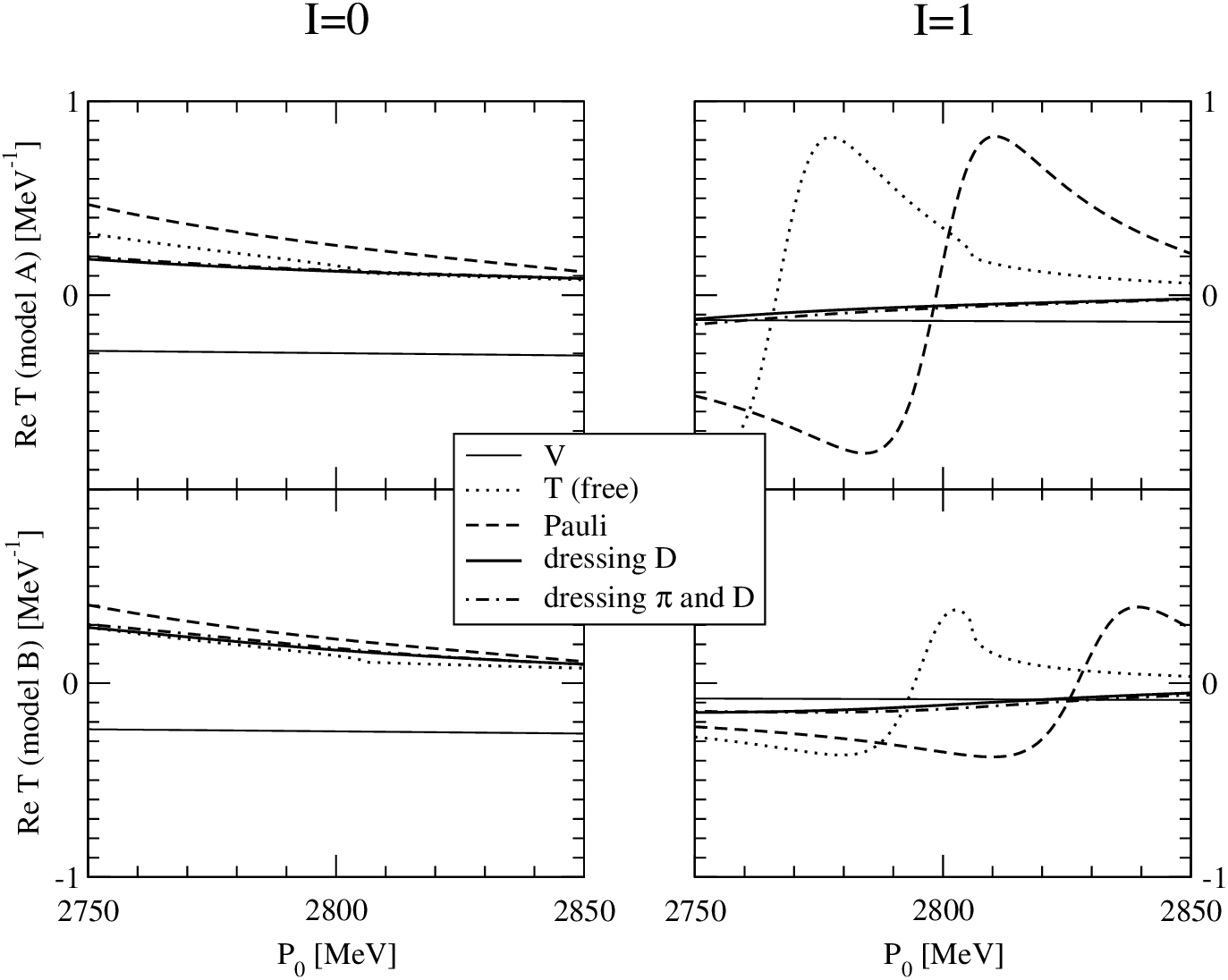}
\caption{Real part of the in-medium $DN$ amplitude for total
momentum $\vec P=0$  as a function of total energy $P_0$ for $I=0$
(left panels) and $I=1$ (right panels) obtained in the cut-off
scheme with model A (upper panels) and model B (lower panels).
The selected range of energies covers values around the $DN$
threshold.} 
\label{Tmed_cut_2}
\end{center}
\end{figure}

In order to illustrate the effect of each approximation on the
sign and strength of the in-medium $DN$ amplitude, we display its
real part in Fig.~\ref{Tmed_cut_2} for
the same cases as in Fig.~\ref{Tmed_cut_1}, but focusing on the
energy region close to the free space $DN$ threshold. The bare interaction
$V$, represented by the thin solid line, is attractive both in the
$I=0$ and $I=1$ channels. The dotted line represents the free $DN$
amplitude, for which we see that the repulsive effect induced by
the $I=0$
$\tilde\Lambda_c(2593)$ is still visible at energies around 2800
MeV. This effect is more pronounced for the $I=1$ channels because
the $\tilde\Sigma_c(2770)$ resonance is just below the $DN$ threshold.
When Pauli blocking effects are included (dashed lines), we observe
the same qualitative behavior except that
the amplitudes are shifted to higher energies.
The fully self-consistent
amplitudes, both with pion dressing (dot-dashed lines) or without (solid lines)
show similar features in the $I=0$
channels as in the other approximations. But in the $I=1$ channel a drastic
dilution of the $\tilde\Sigma_c(2770)$ resonance ends up producing
a mildly attractive
interaction which partly compensates the repulsion found in the $I=0$ sector.

\begin{figure}[t]
\begin{center}
\includegraphics[width=14cm]{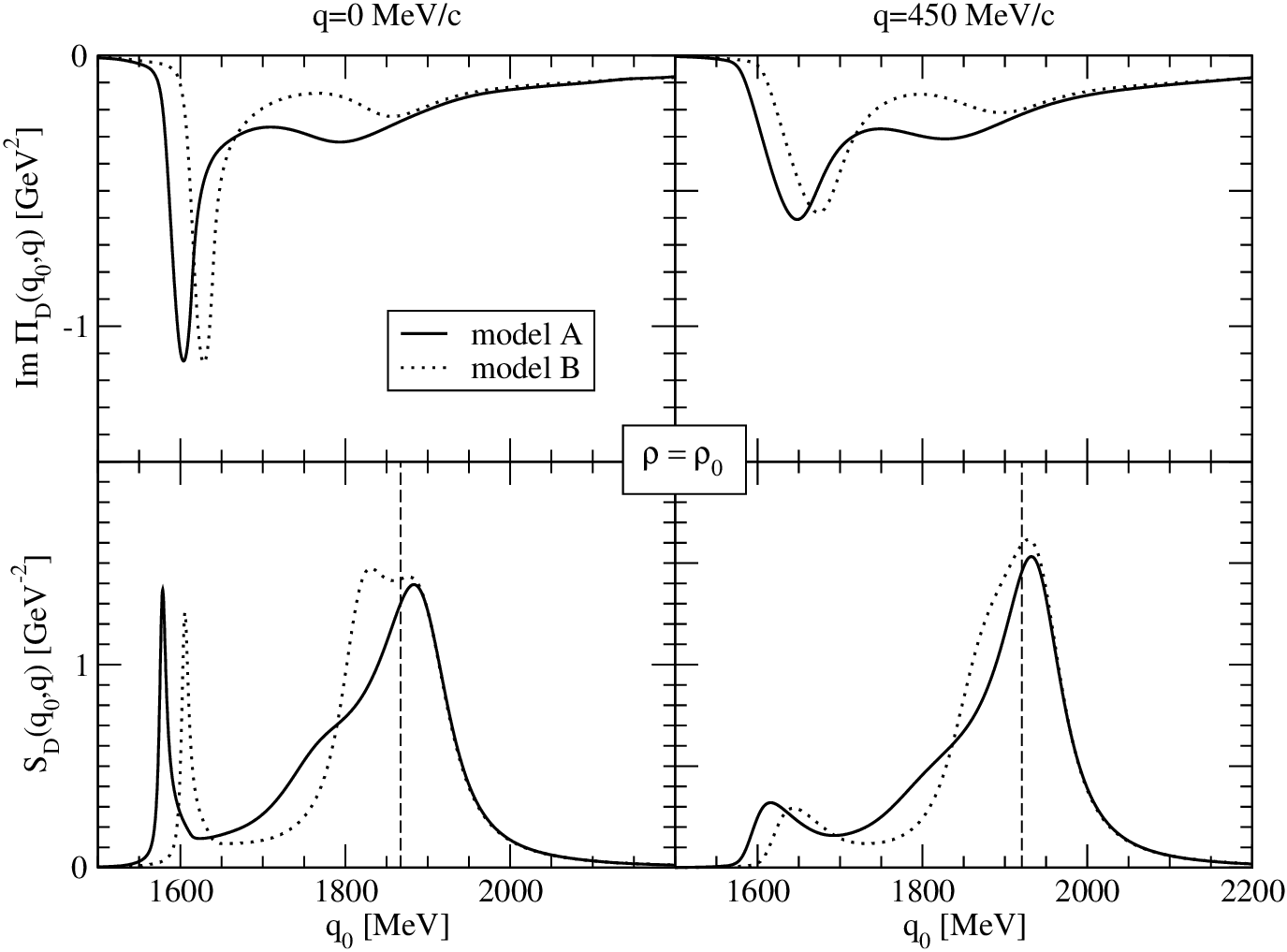}
\caption{Imaginary part of the $D$ meson self-energy (upper
panels) and corresponding $D$ meson spectral function (lower
panels) as functions of energy for two values of the meson
momentum, $q=0$ and $q=450$ MeV at normal nuclear matter density.
Results are shown for model A (solid lines) and model B (dotted
lines). Thin vertical lines display the location of the
$D$-meson pole in free space.} 
\label{spec_cut_2}
\end{center}
\end{figure}

Finally, in Figs.~\ref{spec_cut_2} and 8 we
present our results for the imaginary part of the $D$ meson
self-energy (upper panels), as well as the corresponding spectral
function (lower panels),  as functions of the meson energy $q_0$
for nuclear matter densities $\rho=\rho_0$ and $2\rho_0$,
respectively. We show results for two values of the meson
momentum, $q=0$ (left panels) and $q=450$ MeV (right panels) and
for models A (solid lines) and B (dotted lines) . The self-energy
presents two peaks, the more pronounced one at lower energy is built up
from $\tilde\Lambda_c N^{-1}$ configurations, while the peak at higher
energy is due to the coupling of the $D$ meson to $\tilde\Sigma_c
N^{-1}$ states. 
%Here $\Lambda_c^*$ and $\Sigma^*_c$ are
%the in-medium analogues of the free space dynamical
%resonances $\Lambda_c(2593)$ and $\Sigma_c(2770)$, respectively.

\begin{figure}[t]
\begin{center}
\includegraphics[width=14cm]{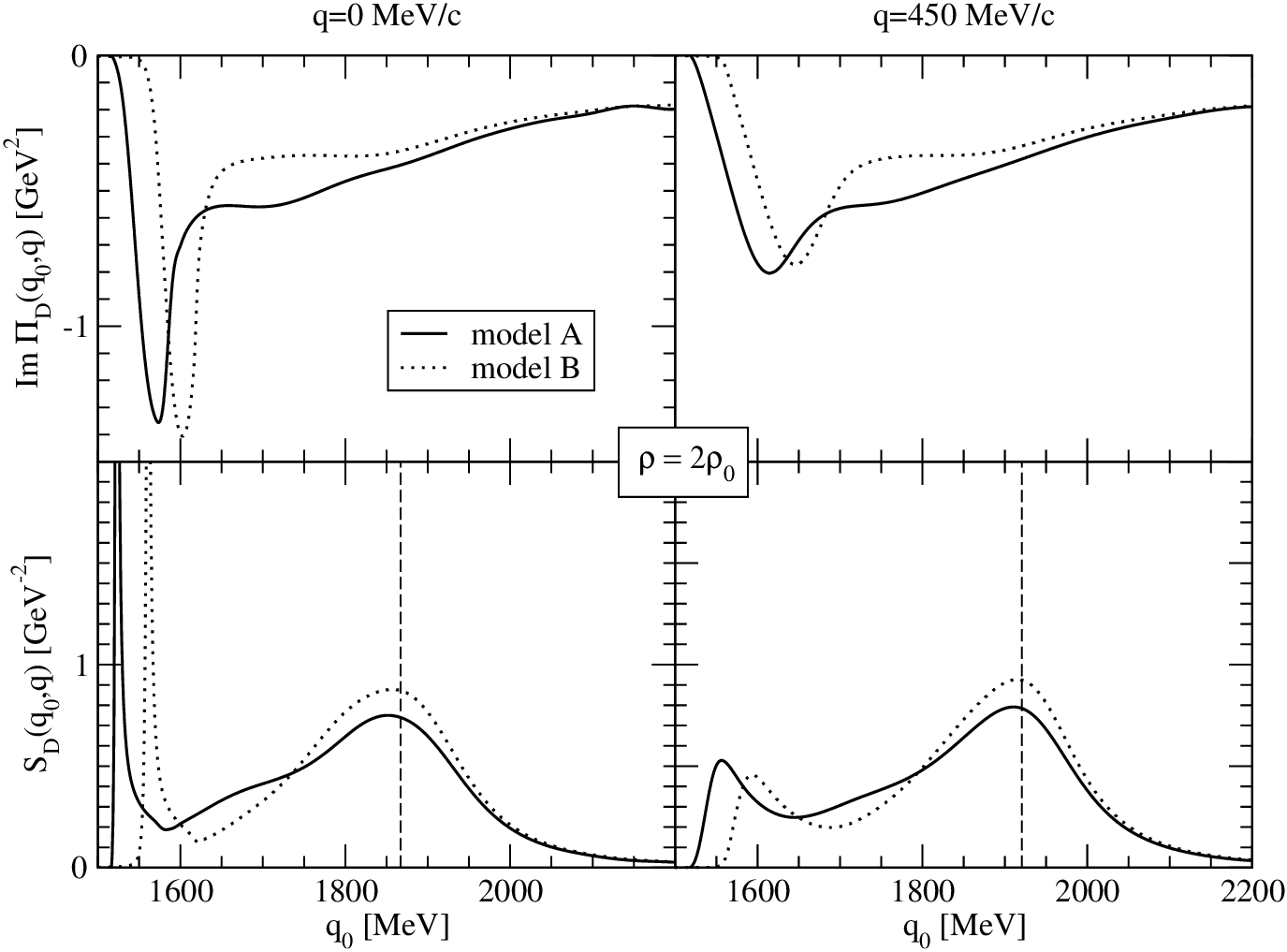}
\caption{The same as Fig.~7 for nuclear
matter density $\rho=2\rho_0$.}
\end{center}
\label{spec_cut_rho2}
\end{figure}

Each peak of
the imaginary part of the $D$-meson self-energy has a direct
association with the structure observed in the spectral function
at a slightly lower energy, as seen in the lower panels. The
narrower peak is the one associated with $\tilde\Lambda_c N^{-1}$
configurations and, for model A, it lies about 50 MeV below what would 
have been the corresponding location if the
$\Lambda_c(2593)$ had not been modified by medium effects.
The in-medium attraction of this lower energy mode 
is more moderate for model B.
By inspecting Fig.~8, we also observe that the amount
of attraction of these $\tilde\Lambda_c N^{-1}$ configurations increases 
by an additional 50 MeV or slightly more when going to twice 
nuclear matter density.
The structure in the spectral function connected to
the bump in the self-energy
associated with the $I=1$ resonance is very faint. It only stands
out in model B and at normal nuclear matter density because of its
proximity to the $D$-meson
quasi-particle peak.

The location of the quasi-particle peak
is mainly driven by the value of quasi-particle energy [Eq.~(\ref{eq:qpe})].
At normal nuclear matter density, the quasi-particle energy was
found to be 5 MeV above the $D$-meson mass. However,
as commented at the beginning of this
section, the actual quasi-particle peak of the spectral function can
appear slightly shifted due the energy dependence of the imaginary part of the
$D$-meson self-energy. At normal nuclear matter density and
for both models A and B,
we find the quasi-particle peak about 20 MeV above
the $D$-meson pole position in free space (denoted by thin
vertical lines in the figures).
At twice nuclear matter density, the quasi-particle peak is
shifted mildly below the free $D$-meson pole in model A, while staying
practically at the same position in model B. 

Our results are qualitatively
similar to those found in \cite{TOL04} but differ considerably from those
in \cite{LUT05}, especially at $2\rho_0$ where the latter reference finds
60 MeV repulsion for the quasi-particle peak of the in-medium $D$-meson. 
This is due to the fact that the $I=1$ resonance is found at a much lower energy than in the present
work, influencing differently the $DN$ interaction around
threshold which becomes substantially more repulsive.
Correspondingly, the contribution of the $I=1$ $DN$ amplitude will affect the
$D$-meson spectral function at lower energies. In fact, 
due to the almost degeneracy between the $I=0$ and $I=1$ charmed resonances found in
the model of \cite{HOF05}, the lower mode of the $D$-meson spectral function in
\cite{LUT05} is dominated by $\tilde\Sigma_c N^{-1}$ configurations
(recall the weight factor 3 for $I=1$ contributions
with respect to $I=0$ ones).

\section{Conclusion}
\label{sec:conclusion}

Now we want to summarize our present work. First, we critically reviewed
several effective hadronic approaches used to investigate the properties of the
$D$ meson in nuclear matter. This then lead us to adopt a recent model by
Hofmann and Lutz: \cite{HOF05}, but with  a few important modifications: (i)
some simplifications in the form of the interactions which have turned out to
be more consistent upon reduction from $t$-channel vector meson exchanges to a
zero range  Tomozawa-Weinberg (T-W) form, (ii) introduction of a supplementary
scalar-isoscalar interaction in the diagonal $DN$ channel, which we call the
$\Sigma_{DN}$ term, apparently prevalent both in the QCD sum rule (QCDSR) and
Mean Field (NMFA) approaches to the problem, and (iii) switching to a
conventional momentum cut-off regularization which was found to be more
consistent than the dimensional method in view of its application to
meson-baryon scattering in nuclear medium. 

In free space, the coupled-channel
equations resulting from the meson-baryon interactions thus obtained were
regularized to reproduce the position and width of the $\Lambda_c(2593)$
resonance in the $I=0$ $DN$ channel. In the $I=1$ channel, the same
interactions were found to generate a wide resonance which we term as
$\Sigma_c(2770)$. 

In nuclear matter, the $DN$ diagonal element of the
interaction was tested in the lowest order approximation to the $D$ meson
self-energy based upon the simplest mean-field
picture,  thus with an equation of the type 
Eq.~(\ref{eq:Kdisp}) for $D$. A $D$ mass reduction of $\sim 60$ MeV was found.
This is consistent with the consequence in QCDSR \cite{HIG00, MOR02} and some 
NMFA results \cite{TSU99, SIB99, MIS104}. If this consistency can be regarded as
important, that would support, at least in part, the introduction of the
$\Sigma_{DN}$ in our scheme, which we have taken to be
quite conservative. It would be interesting to develop models that
could provide a more precise value for this term.

Our in-medium study finds that, 
once the fully self-consistent
coupled-channel equations are solved including Pauli blocking and meson
dressing effects, the situation changes quite drastically. Namely, at
normal nuclear matter density the quasi $D$-meson  peak in the $D$ meson
spectral function is found about $20$ MeV higher than the corresponding
free-space pole position. This appears roughly independent of whether the
$\Sigma_{DN}$ term is present or not. 
The primary cause is found in the fact that the
$\tilde\Sigma_c(2770)$ resonance gets
extremely broadened due to the medium effects.
As the nuclear matter density increases,
the upward shifting of the $D$-meson quasi-particle peak
slows down and can even reverse
if $\Sigma_{DN}\neq 0$. So it may be worth
exploring how this trend continues at even higher densities. 
% or continues
%but very slowly if $\Sigma_{\DN}=0$. 
The $\tilde\Lambda_c(2593)$ resonance keeps narrow and lowers
its position by almost $50$ MeV at normal nuclear matter density
 when the  $\Sigma_{DN}$ term is retained
in the diagonal $DN$ channel. This tendency persists when the matter 
density is doubled. It may be appropriate to stress that, unlike in
the case of the in-medium $\bar K$, the role of the intermediate pion dressing
has been found to be of minor importance here.  This makes a marked difference from 
 the result reported in Ref.~\cite{TOL04} where  the $\tilde\Lambda_c(2593)$ 
 appears to be washed out by the effect of the pion dressing.  

 In the Introduction section, we stated that one of the primary motivations for
studying the behavior of the $D$ meson in nuclear matter is  an attempt to
understand (even partially) the reduction of the $J/\Psi$ charmonium 
production observed in the ultra-relativistic heavy ion reactions, etc.  
%Our eventual objective is no exception to this.  
We are fully aware that the present work 
is just a first step
towards that goal based upon the effective hadronic picture.  In order to be more 
realistic, one of the principal aspects that we will need to investigate is the implementation 
of the finite temperature effect.  As mentioned above,  the finite nuclear density 
makes the $\tilde\Lambda_c(2593)$ 
resonance to move to sensibly lower energies mantaining its narrow width.  
So it should be important 
to study if it may survive the thermal agitation of the order of $\sim 100$ MeV. 
In such a case, the excited charmonia such as 
$\chi_{c\ell}(1P)$ $(\ell=1,2)$, could decay strongly through this 
in-medium resonance, thereby reducing the usual supply of
$J/\Psi$ mesons coming from their radiative decay \cite{ANT93}.
In addition to a realistic implementation of the temperature effect,
one should also have to consider the nuclear mean field binding 
of ground state charmed baryons, a more
extensive study of the scalar attraction characterized 
by the $\Sigma_{DN}$-term, inclusion
of reaction channels with charmed vector mesons, etc. even within the context of 
the effective hadronic picture.

\section{Acknowledgments}

The authors have benefited from various discussions as well as correspondences with many colleagues to whom
they would like to express
their gratitude. Among them are:  M.~Centelles,  M.F.M.~Lutz, A.~Mishra,  
E.~Oset, J.~Schaffner-Bielich, J.~Soto,
A.W.~Thomas, L.~Tol\'os, X.~Vi\~nas and W.~Weise.
Special thanks are due to C.~Garc\'{\i}a-Recio,
C.E. Jim\'enez,
J.~Nieves and L.L.~Salcedoi, and
I. Vida\~na, whose independent study has lead us to locating  a few
small errors in transcribing
our coupling strengths. T.M. is grateful to the support: SAB2004-0135, from the Spanish Ministry of Science
and Research
which enabled him to pay a pleasant visit to  Barcelona.  He is also thankful to the very friendly
environment extended to him
by the members of ECM, Facultat de Fisica, Universitat de Barcelona during the summer and fall, 2005.
This work is partly supported by
contract BFM2003-00856 from
MEC (Spain) and FEDER,
the Generalitat de Catalunya contract 2005SGR-00343,
and the E.U. EURIDICE network contract HPRN-CT-2002-00311.
This research is part of the EU Integrated Infrastructure Initiative
Hadron Physics Project under contract number RII3-CT-2004-506078.

%\section*{Appendix}

%\begin{references}

\end{document}